\documentclass[11pt, a4paper]{article}
\pdfoutput=1

\usepackage{jcappub}

\usepackage{comment}
\usepackage{amsmath}
\usepackage{amsfonts}
\usepackage{amssymb}
\usepackage{graphicx}
\usepackage{mathrsfs}
\usepackage{latexsym}
\usepackage{color}
\usepackage{slashed, cancel}
\usepackage{hyperref}
\usepackage{cleveref}
\usepackage{float}
\usepackage{tikz}
\usepackage{bbold}
\usepackage{soul} 
\usepackage{mathtools}
\usepackage[utf8]{inputenc} 
\usepackage{csquotes}
\usepackage{support-caption} 
\usepackage{subcaption}
\usepackage{enumitem}
\usepackage{orcidlink}
\usepackage{journals}
\allowdisplaybreaks
\usepackage{multirow}
\usepackage{pifont}

\usepackage[version=3]{mhchem}
\hypersetup{urlcolor=blue, colorlinks=true} 

\usepackage{fancyhdr}
\renewcommand{\arraystretch}{1}
\numberwithin{equation}{section}

\definecolor{orange}{rgb}{1,0.4,0}
\definecolor{green}{rgb}{0,0.65,0}
\definecolor{rossos}{rgb}{0.8,0.2,0.3}
\definecolor{bluscuro}{rgb}{0.15, 0.2, .85}
\definecolor{bluchiaro}{cmyk}{1,.3,0.,0.1}
\hypersetup{colorlinks, citecolor=bluscuro, linkcolor=bluscuro, urlcolor=bluscuro}

\makeatother   
\newcommand{\GeV}{{\rm \,GeV}}

\newcommand{\MeV}{{\rm \,MeV}}

\newcommand{\cm}{{\rm \,cm}}

\newcommand{\km}{{\rm \,km}}
\newcommand{\s}{{\rm \,s}}

\newcommand{\K}{{\rm \,K}}

\newcommand{\Gyr}{{\rm \,Gyr}}

\newcommand{\Msun}{M_\odot}
\newcommand{\Mstar}{M_\star}
\newcommand{\Rstar}{R_\star}
\newcommand{\vstar}{v_\star}
\newcommand{\tstar}{t_\star}
\newcommand{\Tstar}{T_\star}

\newcommand{\Teff}{T_{\rm eff}}

\newcommand{\vesc}{v_\mathrm{esc}}

\newcommand{\MsqT}{|\overline{M_T}|^2}

\newcommand{\fMB}{f_{\rm MB}}

\newcommand{\tth}{t_\text{therm}}

\newcommand{\tthD}[1]{\tth^{(D#1)}}

\newcommand{\erf}{{\rm \,Erf}}

\begin{document}

\hfill KCL-PH-TH/2024-20

\hfill INT-PUB-24-016

\hfill FERMILAB-PUB-24-0196-T

\title{Heavy Dark Matter in White Dwarfs: Multiple-Scattering Capture and Thermalization}

\author[a]{Nicole F.\ Bell\,\orcidlink{0000-0002-5805-9828},}
\author[b,c]{Giorgio Busoni\,\orcidlink{0000-0002-8527-0768},}
\author[d,e]{Sandra Robles\,\orcidlink{0000-0002-6046-8217}}
\author[a]{and Michael Virgato\,\orcidlink{0000-0002-8396-0896}}

\affiliation[a]{ARC Centre of Excellence for Dark Matter Particle Physics, \\
School of Physics, The University of Melbourne, Victoria 3010, Australia}
\affiliation[b]{ARC Centre of Excellence for Dark Matter Particle Physics, \\
Research School of Physics,
The Australian National University, Canberra ACT 2601, Australia}
\affiliation[c]{
ARC Centre of Excellence for Dark Matter Particle Physics, \\
Department of Physics, University of Adelaide, South Australia 5005, Australia} 
\affiliation[d]{Theoretical Particle Physics and Cosmology Group, Department of Physics, King’s College London, Strand, London, WC2R 2LS, UK}
\affiliation[e]{Particle Theory Department, Theory Division, Fermi National Accelerator Laboratory, Batavia, Illinois 60510, USA}

\emailAdd{n.bell@unimelb.edu.au}
\emailAdd{giorgio.busoni@adelaide.edu.au}
\emailAdd{srobles@fnal.gov}
\emailAdd{mvirgato@student.unimelb.edu.au}

\abstract{
We present an improved treatment for the scattering of heavy dark matter from the ion constituents of a white dwarf. In the heavy dark matter regime, multiple collisions are required for the dark matter to become gravitationally captured.  Our treatment incorporates all relevant physical effects including the dark matter trajectories, nuclear form factors, and radial profiles for the white dwarf escape velocity and target number densities. Our capture rates differ by orders of magnitude from previous estimates, which have typically used approximations developed for dark matter scattering in the Earth.
We also compute the time for the dark matter to thermalize in the center of the white dwarf, including in-medium effects such as phonon emission and absorption from the ionic lattice in the case where the star has a crystallized core. We find much shorter thermalization timescales than previously estimated, especially if the white dwarf core has crystallized.
We illustrate the importance of our improved approach by determining the cross section required for accumulated asymmetric dark matter to self-gravitate.
}

\maketitle

\section{Introduction}

White dwarfs (WDs) are the most abundant stellar remnants; in fact more than $90\%$ of the stars in the Galaxy are WDs. Due to their high density, extreme conditions and the existence of observational data, they have recently been used to test and constrain dark matter (DM) models. This generally involves the accumulation of DM particles, which leads to observational signatures such as an increase in their luminosity~\cite{Bertone:2007ae,McCullough:2010ai,Hooper:2010es,Amaro-Seoane:2015uny,Dasgupta:2019juq,Panotopoulos:2020kuo,Curtin:2020tkm,Bell:2021fye} 
or DM-triggered supernova ignition/black hole formation~\cite{Graham:2015apa,Bramante:2015cua,Graham:2018efk,Acevedo:2019gre,Janish:2019nkk,Steigerwald:2022pjo}. The latter scenario specifically requires the capture of heavy DM of order 100 TeV, or above. For these masses, multiple collisions of the DM with the WD constituents are required for the DM to lose sufficient energy to become gravitationally bound to the star.

Analytical approaches in the literature that deal with capture of heavy DM via multiple scattering in stars \citep{Bramante:2017xlb,Dasgupta:2019juq,Ilie:2020vec,Ilie:2021iyh} are based on Gould's seminal work for capture in the Earth \citep{Gould:1991va}. It is worth noting that in the case of the Earth, the main contribution comes from the scattering of DM with iron nuclei. 
Gould's formalism was derived under the following assumptions: (i)  DM trajectories are unaffected by collisions, 
(ii) constant escape velocity, 
(iii) constant iron density, 
(iv) DM follows linear trajectories outside and inside the  Earth's core, thereby neglecting gravitational focusing. 
All of these assumptions approximately hold in the Earth's core, where the iron density profile is rather flat, and the  gravitational potential of the Earth is weak, meaning that the DM escape velocity is much smaller than the average DM halo speed. However, in the case of the Earth, a Monte Carlo simulation is in fact best suited to tackle multi-scattering capture of light DM~\cite{Bramante:2022pmn}.

For WDs, on the other hand, we have a very different kinematic regime. This enables an analytical approach to be attempted. For capture in WDs, the infalling DM particles are accelerated to velocities that are a sizeable fraction of the speed of light and hence orders of magnitude greater than the DM halo speed. In this regime, the assumptions (ii)--(iv) above do not hold. (Assumption~(i) remains well justified.) 
In this paper, we relax all the non-valid simplifying assumptions, and derive a sound formalism for multiple scattering capture in WDs.  We show that the simplifying assumptions would lead to an overestimation of the DM capture rate, with the difference becoming more severe for heavier DM. 

Our formalism is based on a response function that properly encodes the relevant physical effects, including the cross section suppression due to the nuclear form factors; non-constant escape velocity; and gravitational focusing when calculating the probability for DM to undergo more than one collision and lose enough energy in the process to become bound to the WD. We shall begin by considering DM scattering from a single type of WD constituent. 
As we shall see, this method can be generalized to the more realistic case of WDs made of more than one ionic species.

Following gravitational capture, the DM particles will continue to scatter with the WD constituents until an equilibrium state is reached. The timescale for this process to occur has previously been estimated~\cite{Acevedo:2019gre,Janish:2019nkk,Steigerwald:2022pjo} by using the average DM energy loss in an energy regime that is suited for capture, i.e. for high energy transfers. We show that due to finite temperature effects the average energy loss has a different scaling with the DM kinetic energy and DM mass for high and low energy momentum transfers. The latter is more important for thermalization. In addition, if the WD is old enough that its core has begun to crystallize, then the ions will form a Coulomb lattice, in particular in the region near the center of the star. In this case, collective excitations of the ionic targets introduce an additional correction which we find to increase the thermalization time by less than an order of magnitude, much less than previously estimated.

Finally, to showcase the effect of using our revised calculations for capture and thermalization in the multiple-scattering regime, we consider the accumulation of heavy non-annihilating dark matter.  Such non-annihilating dark matter is expected in asymmetric dark matter models, where the DM abundance in the present-day Universe consists entirely of DM particles and no DM antiparticles (or vice-versa). 
We shall use our formalism to estimate the size of the DM-nucleon cross section required for non-annihilating DM accumulated in the WD core to self-gravitate. Since we are dealing with a DM population that grows over time, the time evolution of the WD core temperature and hence that of thermalization time is considered. Depending on the composition and compactness of the WD, we find at least one order of magnitude difference with previous results.  

This paper is organized as follows. In section~\ref{sec:wd}, we briefly review the physics of WDs relevant to this work and how to model it in order to use observational data. In section~\ref{sec:capture}, we derive the response function for DM collisions in the multiple-scattering regime, both for scattering from a single target species, and from multiple targets. The timescale for thermalization of heavy DM is discussed in section~\ref{sec:therm}. The condition for self-gravitation of asymmetric DM is examined in section~\ref{sec:SNignition} and our conclusions are given in section~\ref{sec:conclusion}.

\section{White Dwarfs}
\label{sec:wd}

White dwarfs are the most abundant stars in the Galaxy, being the end points of main sequence stars with mass $\lesssim 8-10 \; M_\odot$, depending on metallicity. This compact stellar remnant is supported from gravitational collapse by electron degeneracy pressure. WDs are born at high temperatures, $\sim 10^8\K$, and cool over billions of years to $\sim 10^5\K$. 
Below we discuss how the internal structure is modeled from an appropriate equation of state, and summarize the observational data used in this work.

\subsection{Evolution and Internal Structure}
\label{sec:wdsimus}

WDs are born  as extremely hot, compact objects, whose fate is to cool slowly by radiating light. There is no longer nuclear burning in the core of these collapsed objects. The only energy source, responsible for the star's luminosity, is of thermal origin: the kinetic motion of the non-degenerate ions which are completely decoupled from the electrons. The degenerate electrons do not contribute to the cooling process. They determine the star's stellar structure and hence the WD mass radius relation. In addition, as excellent heat conductors, electrons are very efficient at thermalizing the WD core, and hence we can model this region of the WD as an isothermal sphere that accounts for $\sim99\%$ of the star's mass. 
As the WD evolves (cools), its ionic constituents undergo phase transitions from a hot gas to a liquid and finally to a crystal, i.e an ion lattice. A significant amount of energy is released in the latter first-order phase transition, which slows down the cooling process. The WD solidifies from the center to the surface. 
The remaining $\sim1\%$ of the WD mass is in the form of a heat-blanketing envelope, whose composition, mainly hydrogen and helium, plays a key role in the cooling process.

In ref.~\cite{Bell:2021fye}, we modeled the  WD core as being composed of only one element, namely carbon, by solving the structure equations, the Tolman-Oppenheimer-Volkoff (TOV) equations~\cite{Tolman:1939jz,Oppenheimer:1939ne} coupled to the 
zero-temperature Feynman-Metropolis-Teller (FMT) equation of state (EoS)~\cite{Rotondo:2009cr,Rotondo:2011zz}. 
In the present work, we adopt a more realistic approach to match observations. Specifically, we consider stratified WDs made of two elements, 50\% each, namely carbon-oxygen (CO) for WDs of mass $\lesssim1.05\Msun$, and oxygen-neon (ONe) above this mass threshold. In addition, we use the finite temperature extension of the FMT EoS presented in ref.~\cite{deCarvalho:2013rea}. While the zero-temperature approach is adequate for WDs with masses $\gtrsim 0.7 - 0.8\;M_\odot$, the predicted mass-radius relation in this approximation begins to deviate from observations for lower masses. These deviations are expected to stem from finite temperature effects~\cite{deCarvalho:2013rea}. 
In the relativistic FMT EoS at finite temperature, the thermal energy of the nuclei is considered when calculating the energy density and pressure within a Wigner-Seitz cell\footnote{A Wigner-Seitz cell encloses a cloud of electrons surrounding a finite size nucleus so that the system is neutral.}. The electron energy density and pressure depend on the degeneracy parameter and ultimately on the temperature of the system; for further details see ref.~\cite{deCarvalho:2013rea}. EoSs for C, O and Ne were obtained and later used together with the TOV equations to obtain WD radial profiles.
This includes the radial profiles of the escape velocity, $\vesc$, an example of which can be found in Fig.~1 of ref.~\cite{Bell:2021fye}.

\subsection{Observations}

\begin{table}[t]
    \centering
    \setlength{\tabcolsep}{0.275em}   
    \begin{tabular}{|l|c|c|c|c|c|c|}
    \hline
     \multirow{2}{*}{WD name} &   $\Teff$ & 
     $\Mstar$ &  Cool.\,age 
    & Distance &  $\Tstar(\tstar)$  & Onset \\ & (K) & $(\Msun)$ & $\tstar$ (Gyr) & (pc) & $(10^6 \, \K)$ & crys.$\,(\Gyr)$  \\
    \hline
      WD 0821-669 (CO) & 4808\,\cite{Limbach:2022} &  
      0.53\,\cite{Limbach:2022}   & 6.58\,\cite{Limbach:2022}  &  10.68 \cite{GaiaDR3:2020} & $1.8$  &  3.09\\
      GJ 3182 (CO) & 4980\,\cite{Limbach:2022} &  
      0.62\,\cite{Limbach:2022} &  7.24\,\cite{Limbach:2022} & 10.87 \cite{GaiaDR3:2020} & $1.9$ & 2.36 \\
     SDSS J232257.27+252807.42 (ONe) & 6190\,\cite{Camisassa:2019} & 
     1.11\,\cite{Camisassa:2019} & 4.58\,\cite{Camisassa:2019} 
     &199.17\,\cite{GaiaDR3:2020} &  $2.1$  & 2.73\\            \hline 
    \end{tabular}
    \caption{
    Old WDs within 200 pc with spectral type DA. 
    The first 3 WD properties, namely cooling ages, effective surface temperatures and masses for the last two WDs are given in refs.~\cite{Camisassa:2019,Limbach:2022}, while the last two columns, core temperature at the cooling age and onset of crystallization, were  obtained by using the evolutionary sequences of refs.~\cite{Bedard:2020} and \cite{Camisassa:2019} for carbon-oxygen and oxygen-neon  WDs, respectively. 
    }
    \label{tab:localWDs}
\end{table}

Over the past few decades, our understanding of white dwarfs has greatly been improved with the aid of large-area surveys, in particular the Sloan Digital Sky Survey (SDSS)~\cite{York:2000gk} which has produced the largest catalogue of spectroscopically confirmed WDs, see e.g. refs.~\cite{Kleinman:2013,Kepler:2015,Kepler:2019}.
Spectroscopy provides measurements of the effective temperature $\Teff$ and surface gravity $g$ of a WD, from which the WD mass $\Mstar$ and radius $\Rstar$ can be inferred.  
Spectroscopic data also give us information about the composition of the atmosphere of a white dwarf (spectral type). These outer layers are key to determining the core temperature $\Tstar$, cooling age and other WD properties, with the help of model atmospheres.

\begin{figure}
    \centering
    \includegraphics[width=0.495\textwidth]{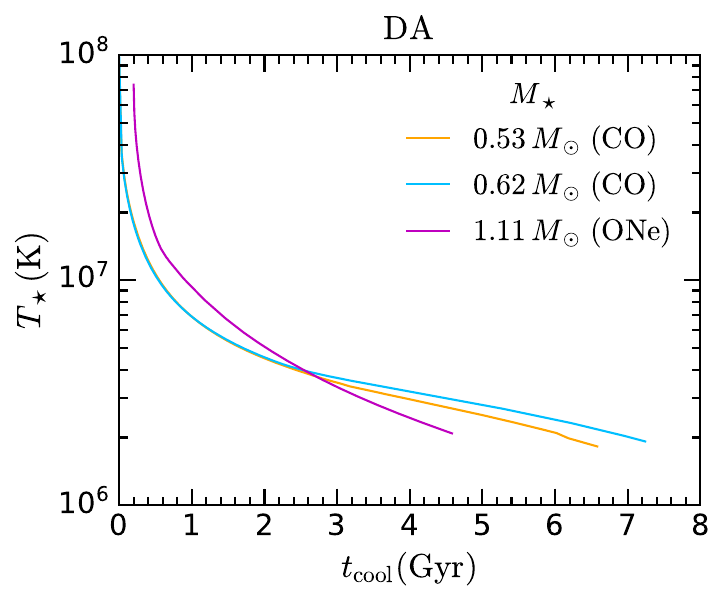}
    \includegraphics[width=0.495\textwidth]{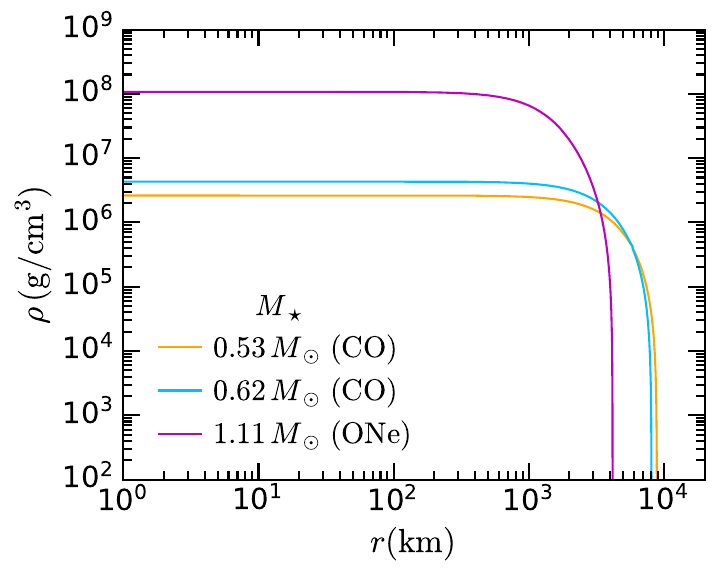}    
    \caption{Left: Evolution of the core temperature $\Tstar$ for the WDs  in Table~\ref{tab:localWDs}.
    Right: Radial profiles of the core density for the same WDs  with core temperatures listed in Table~\ref{tab:localWDs}.}
    \label{fig:Tstarevol}
\end{figure}

In Table~\ref{tab:localWDs}, we show three old WDs  within 200 pc from the Sun, distances measured by Gaia EDR3~\cite{GaiaDR3:2020}. These WDs have spectral type DA, i.e. a hydrogen rich atmosphere. 
In order to model these stars, rather than simulating their full evolution, one can track how the core temperature and other relevant properties change over time by using evolutionary tracks based on model atmospheres such as those given in ref.~\cite{Bedard:2020} for CO WDs and ref.~\cite{Camisassa:2019} for ONe WDs~\footnote{Evolutionary sequences for CO WDs of various masses can be found at: \url{https://www.astro.umontreal.ca/~bergeron/CoolingModels/} and for ONe WDs at \url{http://evolgroup.fcaglp.unlp.edu.ar/TRACKS/ultramassive.html}.}. 
The most massive WDs are expected to be made of oxygen and neon, hence their effective surface temperature, mass, core temperature ($\Tstar$), cooling age and age at the onset of core crystallization have been calculated using the evolutionary tracks for massive WDs of ref.~\cite{Camisassa:2019}. The remaining less massive WDs, on the other hand, are expected to be a mixture of carbon and oxygen,  so that their corresponding inferred properties were estimated using the cooling models of ref.~\cite{Bedard:2020}. The evolution of the core temperature as a result of the cooling processes  for the WDs in Table~\ref{tab:localWDs} is depicted in the left panel of Fig.~\ref{fig:Tstarevol}. On the right panel of this figure, we show the radial density profiles of the WD core for the same stars, obtained  as outlined in section~\ref{sec:wdsimus},  using the finite temperature FMT EoS for the core temperature  at the WD cooling age $\tstar$ listed in Table~\ref{tab:localWDs}.

\section{Multiple Scattering Capture}
\label{sec:capture}

In this section, we review the formalism for single-scattering capture of DM in white dwarfs. Building upon it, we derive a proper treatment for multiple-scattering capture, by including the radial dependence of the escape velocity and target number density, the effect of the nuclear form factors,  
as well as considering that DM trajectories follow the rules of gravity. 
Next, we generalize this formalism to the case of multiple targets.

\subsection{Single Scattering}
\label{sec:singlescatt}

We begin by considering the situation where a single collision with the non-relativistic, non-degenerate ions in the WD core is enough for DM to become gravitationally bound to the star. 
In this case, the most general expression for the DM capture rate   $C_1$ in the optical thin limit is given by~\cite{Bell:2021fye}
\begin{equation}
     C_{1} =
\frac{\rho_\chi}{m_\chi}\int_0^{R_\star} dr 4\pi r^2
     \int_0^\infty du_\chi \frac{w(r)}{u_\chi} \fMB(u_\chi) \Omega_T^-(w),\label{eq:ioncapdef}
\end{equation}
with
\begin{eqnarray}   
     \Omega_T^-(w) &=& \int_0^{v_\mathrm{esc}(r)}dv \, R_T^-(w\to v),\label{eq:iongammadef}\\     
    R_T^-(w\rightarrow v) &= &  \int_0^\infty ds\; \int_0^\infty dt  \frac{32 \mu_+^4}{\sqrt{\pi}} k^3 n_T(r) \frac{d\sigma_{T\chi}}{d\cos\theta_\mathrm{cm}} \frac{v t}{w} e^{-k^2 u_T^2} \Theta(t + s - w) \Theta(v - |t - s|),\label{eq:iondiffintrate}
\end{eqnarray}
where $\rho_\chi$ is the DM density, $m_\chi$ is the DM mass, $n_T$ is the target number density,  and $w(r)=\sqrt{u_\chi^2+v_\mathrm{esc}^2(r)}$ and $v$ are the DM velocity before and after the collision, respectively. Kinematic quantities are defined in the center of mass (cm) frame, where $\theta_\mathrm{cm}$ is the center of mass angle, $\sigma_{T\chi}$ is the DM-ion cross section,  
$s$ is the cm speed, $t$ is the velocity of the DM in the cm frame prior to the interaction, $k^2 = m_T/2\Tstar$,  $m_T $ is the mass of the target and $u_T$ is the speed of the target before the collision  in the cm frame: 
\begin{eqnarray}
    u_T^2  &=& 2 \mu \mu_+ t^2 + 2 \mu_+ s^2 - \mu w^2,\label{eq:cmvel}\\
    \mu &=& \frac{m_\chi}{m_T},\qquad 
    \mu_\pm = \frac{\mu\pm 1}{2}.     
\end{eqnarray}

We assume that the DM  speed distribution  in Eq.~\ref{eq:ioncapdef}, $u_\chi$, follows a Maxwell-Boltzmann (MB) distribution, which reads~\cite{Busoni:2017mhe} 
\begin{equation}
\fMB(u_\chi)du_\chi=\frac{u_\chi}{v_dv_\star}\sqrt{\frac{3}{2\pi}}
\left(\exp\left[-\frac{3}{2v_d^2}\left(u_\chi-v_\star\right)^2\right]-\exp\left[-\frac{3}{2v_d^2}\left(u_\chi+v_\star\right)^2\right]\right)du_\chi, 
\label{eq:distribution_relative_distribution}
\end{equation}
where  $v_d$ is the velocity dispersion of the halo and $\vstar$ is the WD velocity in the Galactic rest frame. 
 In the $T_\star\rightarrow0$ limit the interaction rate $\Omega^-$,  Eq.~\ref{eq:iongammadef}, becomes \cite{Busoni:2017mhe}
\begin{equation}
\Omega_T^-(w) = \frac{4\mu_+^2}{\mu w}n_T(r)\int_{w(r)\frac{|\mu_-|}{\mu_+}}^{\vesc(r)}dv v  \frac{d\sigma_{T\chi}}{d\cos\theta_\mathrm{cm}} (w,q_\mathrm{tr}^2(w,v)),\label{eq:intrateions}
\end{equation}
where $q_\mathrm{tr}$ is the momentum transferred during the collision. 
Note that the differential cross section depends on the DM initial velocity $w$ and the momentum transfer through the form factors $F_{i,j}^{(N,N')}$.  We express the cross section as 
\begin{eqnarray}
\frac{d\sigma_{T\chi}}{d\cos\theta_\mathrm{cm}}(w,q_\mathrm{tr}^2)&=&\frac{1}{32\pi}\frac{\MsqT(w,q_\mathrm{tr}^2)}{(m_\chi+m_T)^2}, \label{eq:diffxsec} \\     
\MsqT(w,q_\mathrm{tr}^2) &=& \frac{m_T^2}{m_N^2}\sum_{i,j} \sum_{N,N'} {\cal C}_i^N {\cal C}_j^{N'} F_{i,j}^{(N,N')}(w,q_\mathrm{tr}^2),
\end{eqnarray}
where $N=\{p,n\}$, the spin averaged squared matrix elements are given in the basis of  non-relativistic (NR) operators~\cite{DelNobile:2013sia}, and the ${\cal C}_i$ are the coefficients that accompany the corresponding NR operators as given in Table~\ref{tab:operatorsle}.
The nuclear response functions
$F_{i,j}^{(N,N')}(w,q_\mathrm{tr}^2)$ are given in appendix C of ref.~\cite{Catena:2015uha}. 
Note that we have assumed fermionic DM colliding with ion nuclei, with interactions described by the 6-dimensional Effective Field Theory (EFT) operators of Table~\ref{tab:operatorsle}. Note also that since WDs are composed mainly of ionic species with no nuclear spin, namely $^4$He, $^{12}$C, $^{16}$O and $^{20}$Ne, only scattering operators with spin-independent (SI) interactions are relevant.  For these elements in particular, only three response functions are non-zero, namely $W_M, W_{M\Phi^{''}}, W_{\Phi^{''}}$\cite{Fitzpatrick:2012ix,Catena:2015uha}.

At low energy, the cross section for scattering on target nuclei, $\sigma_{T\chi}$, can be related to the SI cross section on a single nucleon, $\sigma_{N\chi}$, using 
\begin{equation}
\frac{d\sigma_{N\chi}}{d\cos\theta_\mathrm{cm}}(w,q_\mathrm{tr}^2)=\frac{1}{32\pi}\frac{1}{(m_\chi+m_N)^2} \sum_{i,j} {\cal C}_i^N {\cal C}_j^{N} \hat{F}_{i,j}^{(N,N)}(w,q_\mathrm{tr}^2),    
\end{equation}
where $\hat{F}$ is the nuclear response function for a nucleon $N$. Thus, in the low energy limit ($w\rightarrow0,q_\mathrm{tr}^2\rightarrow0$) we have 
\begin{equation}
\frac{d\sigma_{T\chi}}{d\cos\theta_\mathrm{cm}}=\frac{(m_\chi+m_N)^2}{(m_\chi+m_T)^2}\frac{m_T^2}{m_N^2} \frac{\sum_{i,j} \sum_{N,N'} {\cal C}_i^N {\cal C}_j^{N'} F_{i,j}^{(N,N')}(w\rightarrow0,q_\mathrm{tr}^2\rightarrow0)}{\sum_{i,j} {\cal C}_i^N {\cal C}_j^{N} \hat{F}_{i,j}^{(N,N)}(w\rightarrow0,q_\mathrm{tr}^2\rightarrow0)}\frac{d\sigma_{N\chi}}{d\cos\theta_\mathrm{cm}}. 
\end{equation} 
For operators D1, D2, D5 and D6, for which only the $W_M$ response function is present, this simplifies to 
\begin{equation}
\frac{d\sigma_{T\chi}}{d\cos\theta_\mathrm{cm}}=\frac{(m_\chi+m_N)^2}{(m_\chi+m_T)^2}\frac{m_T^2}{m_N^2} A^2 \frac{d\sigma_{N\chi}}{d\cos\theta_\mathrm{cm}}\rightarrow  A^4 \frac{d\sigma_{N\chi}}{d\cos\theta_\mathrm{cm}}, \label{eq:tarnucleonxsec}
\end{equation}
where the final expression is obtained by taking the large $m_\chi$ limit. For D10, the precise coefficient that multiplies the DM-nucleon cross section in the RH expression of Eq.~\ref{eq:tarnucleonxsec} depends on the target, being $\gamma A^4$ with $\gamma=1$ for He, $\gamma=1.56$ for C, $\gamma=1.02$ for O and $\gamma=1.22$ for Ne. 

\begin{table}
\centering
{ \renewcommand{\arraystretch}{1.3}
\begin{tabular}{ | c | c | c | c |}
  \hline                        
  Name & Operator & Coupling & $M_N^{\rm NR} = {\cal C}_i^N {\cal O}_i^{{\rm NR}}$ \\   \hline
  D1 & $\bar\chi  \chi\;\bar N  N $ & ${c_N^S}/{\Lambda^2}$ &  $ 4\frac{ic_N^S}{\Lambda^2} m_\chi m_N  {\cal O}_1^{\rm NR}$ \\  \hline  
  D2 & $\bar\chi \gamma^5 \chi\;\bar N N $ & $i{c_N^S}/{\Lambda^2}$ &  $-4 \frac{ic_N^S}{\Lambda^2} m_N {\cal O}_{11}^{\rm NR}$  \\  \hline  
  D5 & $\bar \chi \gamma_\mu \chi\; \bar N \gamma^\mu N$ & ${c_N^V}/{\Lambda^2}$ & $4 \frac{c_N^V}{\Lambda^2} m_\chi m_N {\cal O}_1^{\rm NR}$\\  \hline   
  D6 & $\bar\chi \gamma_\mu \gamma^5 \chi\; \bar  N \gamma^\mu N $ & ${c_N^V}/{\Lambda^2}$ &   $ 8  \frac{c_N^V}{\Lambda^2} (m_\chi m_N  {\cal O}_8^{\rm NR} + m_\chi {\cal O}_9^{\rm NR}) $ \\  \hline   
 D10 & $\bar \chi \sigma_{\mu\nu} \gamma^5\chi\; \bar N \sigma^{\mu\nu} N \;$ & $i{c_N^T}/{\Lambda^2}$ &  $ 8 \frac{ic_N^T}{\Lambda^2} ( m_\chi {\cal O}_{11}^{\rm NR} - m_N{\cal O}_{10}^{\rm NR} -4 m_\chi m_N{\cal O}_{12}^{\rm NR}) $   \\  \hline
\end{tabular}}
\caption{Relationship between high energy EFT scattering operators (second column) and the non-relativistic (NR) operators $M_N^{\rm NR} = {\cal C}_i^N {\cal O}_i^{{\rm NR}}$~\cite{DelNobile:2013sia} (fourth column) for spin-independent interactions, where $\Lambda$ is the cutoff scale,  and the coefficients $c_N^S$, $c_N^V$ and $c_N^T$ are the hadronic matrix elements, which can be found in appendix B of ref.~\cite{Bell:2021fye}. 
\label{tab:operatorsle}}
\end{table}

In most cases, for  WDs in the solar neighborhood, we have $\vesc(r)\gg u_\chi$ and hence $w(r)\sim \vesc(r)$. Using this approximation, we can rewrite Eqs.~\ref{eq:ioncapdef} and \ref{eq:intrateions} in terms of the recoil energy of the target $E_R$ 
\begin{equation}
    E_R = \frac{q_\mathrm{tr}^2}{2m_T}, \qquad  E_R = E_R^\mathrm{max}\left(\frac{1-\cos\theta_\mathrm{cm}}{2}\right), 
\end{equation}
where
\begin{eqnarray}
    E_R^\mathrm{max} &=& 2m_T \vesc(r)^2 \left(\frac{m_\chi}{m_T+m_\chi}\right)^2, \label{eq:ER_max} \\
    E_R^\mathrm{min}&=&0.
\end{eqnarray}
Trading $\theta_\mathrm{cm}$ and the DM final speed $v$ for the recoil energy $E_R$, we obtain
\begin{eqnarray}
    \frac{d\sigma_{T\chi}}{d\cos\theta_\mathrm{cm}} &=& \frac{E_R^\mathrm{max}}{2}  \frac{d\sigma_{T\chi}}{dE_R}, \\
     \Omega_T^-(w) 
    &=& \frac{4\mu_+^2}{\mu w}n_T(r) \frac{E_R^\mathrm{max}(\vesc,m_\chi,m_T)}{2m_\chi} \int_{E_R^\mathrm{min}}^{E_R^\mathrm{max}}dE_R \frac{d\sigma_{T\chi}}{dE_R}(\vesc,E_R). 
\end{eqnarray}

In the original derivation of Eq~\ref{eq:ioncapdef}, an average over the DM angular momentum $J$ (angle of incidence) was performed~\cite{Gould:1987ir}. Here,  we reintroduce this average explicitly since it will become relevant in the optically thick and multiple scattering regimes that depend on the specific trajectory and hence  angle of incidence. Note that this does not affect the result for single scattering in the optically thin limit. Reintroducing the average over $J$  and substituting the last expression into Eq.~\ref{eq:ioncapdef}, leads to 
\begin{equation}
    C_{1} = \frac{\rho_\chi}{m_\chi} \int_0^{R_\star} dr 4\pi r^2 n_T(r) \sigma_{T\chi}(\vesc(r)) \vesc^2(r) \int_0^1 \frac{ydy}{\sqrt{1-y^2}}  \int_0^\infty du_\chi \frac{\fMB(u_\chi)}{u_\chi} , \label{eq:C1thin}
\end{equation}
where $y=J/J_\mathrm{max}$, and $J_\mathrm{max}$ is the maximum value of $J$ at a given distance from the center of the star. It is worth noting that the integral over the angular momentum $J$ is in fact an integral over all possible DM trajectories, up to isometry transformations (rotations). The factor $y/\sqrt{1-y^2}$ is therefore just a density factor of the number of trajectories with a certain angular momentum corresponding to $y$, and is just due to angular momentum conservation.  Note also that, thanks to the assumption $u_\chi\ll \vesc$, the integral over $u
_\chi$ can be  factored out  from the integral over the volume. 

When the DM-target cross section is larger than a threshold value, the capture rate approaches the geometric (optically thick) limit. In this regime, the flux of DM particles that traverses the star is considerably attenuated, hence the optically thin approximation is no longer appropriate. 
To account for this,  an optical factor $\eta(r)$ is introduced in Eq.~\ref{eq:ioncapdef}, which removes captured DM particles from the incoming flux. This $\eta(r)$ factor,  defined in terms of the optical depth for DM-target  interactions, depends on the trajectory followed by the DM particle until it is captured~\cite{Busoni:2017mhe,Bell:2020jou,Bell:2021fye}. 
There exist two equally probable paths  a DM particle can follow to reach a given point $r$ within the WD. Therefore, there are two corresponding optical depths $\tau^-_\chi$ 
(shortest path) and $\tau^+_\chi$ (largest path) given by~\cite{Bell:2020jou,Bell:2021fye} 
\begin{align}
    \tau^-_\chi(r,y)
     &=\int_r^{R_\star}\frac{dx}{\sqrt{1-y^2\frac{J_\mathrm{max}(r)^2}{J_\mathrm{max}(x)^2}}}\frac{\Omega^-(w(x))}{\vesc(x)\sqrt{1-\vesc^2(x)}},\label{eq:tauminusdef}\\
     \tau^+_\chi(r,y)
    &=\int^{r_\mathrm{min}}_r + \int^{\Rstar}_{r_\mathrm{min}} \frac{dx}{\sqrt{1-y^2\frac{J_\mathrm{max}(r)^2}{J_\mathrm{max}(x)^2}}}\frac{\Omega^-(w(x))}{\vesc(x)\sqrt{1-\vesc^2(x)}} = 2\tau^-_\chi(r_\mathrm{min},y)-\tau^-_\chi(r,y),\label{eq:tauplusdef}
\end{align}
where $r_\mathrm{min}$ is the position of the perihelion. 
For a depiction of the two possible trajectories see Fig.~11 of ref.~\cite{Bell:2020jou}, and 
for details on the derivation of these expressions, see Section 4.2 and Appendix C.1 of ref.~\cite{Bell:2020jou}. 
Then, the optical factor averaged over the two possible trajectories and the DM angular momentum is 
\begin{equation}
    \eta(r)=\frac{1}{2}\int_0^1\frac{y dy}{\sqrt{1-y^2}}\left(e^{-\tau_{-}(r,y)}+e^{-\tau_{+}(r,y)}\right).\label{eq:eta}
\end{equation}
Finally introducing this factor in Eq.~\ref{eq:C1thin}, the full expression for the capture rate in the single scattering regime reads 
\begin{equation}
    C_{1} = \frac{\rho_\chi}{m_\chi} \int_0^\infty du_\chi \frac{\fMB(u_\chi) }{u_\chi}\int_0^{R_\star} dr 4\pi r^2 n_T(r) \vesc^2(r) \sigma_{T\chi}(\vesc(r)) \eta(r). \label{eq:C1opacity}
\end{equation}
For large DM masses, however, this expression will overestimate the true capture rate. 
 This is because it assumes that a single collision is enough to capture a DM particle, consistent with the use of $E_R^\mathrm{min}=0$. In the next section, we go beyond this limitation by developing a formalism to handle capture in the multiple-scattering regime. 

\subsection{Response Function for Multiple Scattering}
\label{sec:msrespfunct}

When the DM is heavy enough, Eq.~\ref{eq:C1opacity} (which assumes a capture probability $\sim1$) would overestimate the correct result. In this regime, more than one collision is necessary for the DM to be captured.  
To account for this, we construct a function that incorporates both of the non-linear effects into the capture rate, namely the star opacity and the multiple-interactions with the ionic target species.

First, we define a probability density function for the energy lost by a DM particle in a collision as 
\begin{equation}
    f(E_R)= \frac{1}{\sigma_{T\chi}}\frac{d\sigma_{T\chi}}{dE_R}(E_R).    
\end{equation}
Then, the probability that the DM loses an amount of energy of at least $\delta E =m_\chi u_\chi^2/2$ after a  single collision is given by
\begin{equation}
     {\cal F}_1(\delta E)=\int_{\delta E}^{\infty}dE_R f(E_R),
      \label{eq:elossscat} 
\end{equation}
where we have extended the upper integration limit from $E_R^\mathrm{max}$ to infinity. This is valid for white dwarfs, as the escape velocity is very large. 
In the same way, the probability of losing the same amount of energy after N scatterings is 
\begin{equation}
     {\cal F}_N(\delta E)=\int_{0}^{\delta E}dE_R {\cal F}_{N-1}(\delta E-E_R)f(E_R).
 \label{eq:elossNscat}     
\end{equation}

For simplicity, we assume that the DM-target cross section is  well approximated by 
\begin{equation}
 \frac{d\sigma_{T\chi}}{d\cos\theta_\mathrm{cm}}\propto e^{-\frac{E_R}{E_0}},
 \label{eq:diffxsecprop}
\end{equation} 
where $E_R$ is the recoil energy and $E_0$ depends on the specific nuclear target. 
That is, we assume exponential nuclear form factors similar to the Helm approximation.   
This leads to 
\begin{eqnarray}
       f(E_R)&=&\frac{\Theta(E_R)}{E_0}e^{-\frac{E_R}{E_0}},\label{eq:probdensfunc}\\
        {\cal F}_1(\delta E) &=& e^{-\frac{\delta E}{E_0}}.
\end{eqnarray}
Defining the dimensionless quantity
\begin{equation}
    \delta = \frac{\delta E}{E_0}=\frac{m_\chi u_\chi^2}{2 E_0}, 
\end{equation}
and taking the  Laplace transform of the $\cal F$ functions written in terms of $\delta$, we find 
\begin{equation}
    \tilde{\cal F}_1(s)=\frac{1}{1+s}, \qquad
    \tilde{\cal F}_N(s)=\frac{1}{(1+s)^N},     
\end{equation}
where the last expression corresponds to 
\begin{equation}
    {\cal F}_N(\delta) = \frac{e^{-\delta}\delta^{N-1}}{N-1!}. 
\end{equation}

To account for multiple collisions, we define the probability that the DM undergoes $N$ scatterings with nuclei. For DM passing through the same path length $\tau_\chi$, this is given by the Poisson distribution~\cite{Gould:1991va}
 \begin{equation}
    p_N(\tau_\chi) = e^{-\tau_\chi}\frac{\tau_\chi^N}{N!}.
    \label{eq:pNtau}
\end{equation}
Substituting ${\cal F}_1(\delta)$ and $p_0$ into Eq.~\ref{eq:C1thin}, we obtain for single scattering
 \begin{equation}
    C_{1} = \frac{\rho_\chi}{m_\chi}\int_0^{R_\star} dr 4\pi r^2 n_T(r) \sigma_{T\chi}(\vesc(r)) \vesc^2(r)\int_0^1 \frac{ydy}{\sqrt{1-y^2}} \int_0^\infty du_\chi \frac{\fMB(u_\chi) }{u_\chi}p_0(\tau_\chi){\cal F}_1(\delta).
\end{equation}
This is equivalent to Eq.~\ref{eq:C1opacity}, which accounts for the star opacity, once we average over the two equally probable optical depths, $\tau_\chi^+$ and $\tau^-_\chi$. Note that ${\cal F}_1$ indicates the probability for DM to be captured after a single collision.
We can now generalize the above expression to the case of $N$ scatterings as
 \begin{equation}
   C_{N} = \frac{\rho_\chi}{m_\chi}\int_0^{R_\star} dr 4\pi r^2 n_T(r) \sigma_{T\chi}(\vesc(r)) \vesc^2(r)\int_0^1 \frac{ydy}{\sqrt{1-y^2}} \int_0^\infty du_\chi \frac{\fMB(u_\chi)}{u_\chi}p_{N-1} (\tau_\chi) {\cal F}_N(\delta). \label{eq:CN}
\end{equation}
The total capture rate is given by the sum over all $N$ collisions,
\begin{equation}
    C=\sum_N C_N.\label{eq:Ctotdef}
\end{equation}

Next, instead of first evaluating the integrals in Eq.~\ref{eq:CN} and then summing over $N$, we sum the series first by introducing the response function, $G(\tau_\chi,\delta)$
\begin{eqnarray}
    G(\tau_\chi,\delta) &\equiv& \sum_{N=1}^\infty p_{N-1}(\tau_\chi) {\cal F}_N(\delta) = \sum_{N=1}^\infty \frac{e^{-\tau_\chi}\tau_\chi^{N-1}}{(N-1)!}  \frac{e^{-\delta}\delta^{N-1}}{(N-1)!} \nonumber\\
    &=& e^{-\tau_\chi-\delta}I_0\left(2\sqrt{\tau_\chi\delta}\right),\label{eq:respfuncexp}
\end{eqnarray}
where $I_0$ is the  modified Bessel function of the first kind $I_n$ for $n=0$. This response function encodes the probability to lose an amount of energy of at least $\delta E$ through multiple collisions. Therefore, $G(\tau_\chi,\delta)d\tau_\chi$ is the probability that DM collides with an ionic target and loses at least an energy $\delta E$ in the scattering process, after travelling a differential length $\delta\tau_\chi$ in the medium, from a region of the stellar interior with optical depth $\tau_\chi$. 

To account for the two possible optical depths, given by Eqs.~\ref{eq:tauminusdef} and \ref{eq:tauplusdef}, we average over them and over the DM angular momentum, as in section~\ref{sec:singlescatt}. This leads to
\begin{equation}
   \tilde{G}(r,\delta) = \int_0^1 dy \frac{y dy}{\sqrt{1-y^2}} \frac{G(\tau_\chi^-(r,y),\delta)+G(\tau_\chi^+(r,y),\delta)}{2}.\label{eq:respfunctot}
\end{equation}
Note that the response function depends on the radial coordinate through the optical depth and on the DM velocity at infinity, $u_\chi$. Finally, the total capture rate that accounts for the stellar structure, star opacity and multiple scattering reads
\begin{equation}
   C = \frac{\rho_\chi}{m_\chi}\int_0^{R_\star} dr 4\pi r^2 n_T(r) \vesc^2(r) \sigma_{T\chi}(\vesc(r))  \int_0^\infty du_\chi \frac{\fMB(u_\chi) }{u_\chi} \tilde{G}\left(r,\frac{m_\chi u_\chi^2}{2E_0}\right), 
   \label{eq:Ctot}
\end{equation}
where the average over the DM angular momentum has been absorbed in the definition of $\tilde{G}(r,\frac{m_\chi u_\chi^2}{2E_0})$.

Recall that we assumed exponential nuclear form factors to derive Eq.~\ref{eq:Ctot}. 
As in ref.~\cite{Bell:2021fye}, 
we now introduce the nuclear response functions given in appendix~C of ref.~\cite{Catena:2015uha} in Eq.~\ref{eq:diffxsec}. In principle, we can proceed in a similar fashion to the exponential form factor case and derive their corresponding ${\cal F}_N$ functions, as well as the response function for multiple scattering, which is more involved and is given in terms of generalized hypergeometrical functions. A way to circumvent this is to approximate the form factor with an exponential. 
 For instance, if we consider WDs made of carbon, and interactions described by the scalar-scalar operator (D1), we have 
\begin{eqnarray}
   \dfrac{d\sigma_{T\chi}}{dE_R} &\propto& e^{-\frac{E_R}{E_1}}\left(1-x_0\frac{E_R}{E_1}\right)^2, 
\end{eqnarray}
with $x_0\sim0.22$ and $E_1\sim1.25\MeV$, where these factors come from the nuclear response function for $^{12}C$. 
This function is well approximated by $e^{-E_
R/E_0}$
where  $E_0$ is given by 
\begin{equation}
E_0 = \lim_{m_\chi\rightarrow\infty} \dfrac{\int_0^\infty dE_R \frac{d\sigma_{T\chi}}{dE_R} E_R}{\int_0^\infty dE_R \frac{d\sigma_{T\chi}}{dE_R} }.  
\label{eq:E0}
\end{equation}
This  value of $E_0$ is then used in the response function $\tilde{G}$. We have checked that this approximation holds for the spin independent operators D1 and D5 for carbon, oxygen and neon targets.

\begin{figure}
    \centering
    \includegraphics[width=\textwidth]{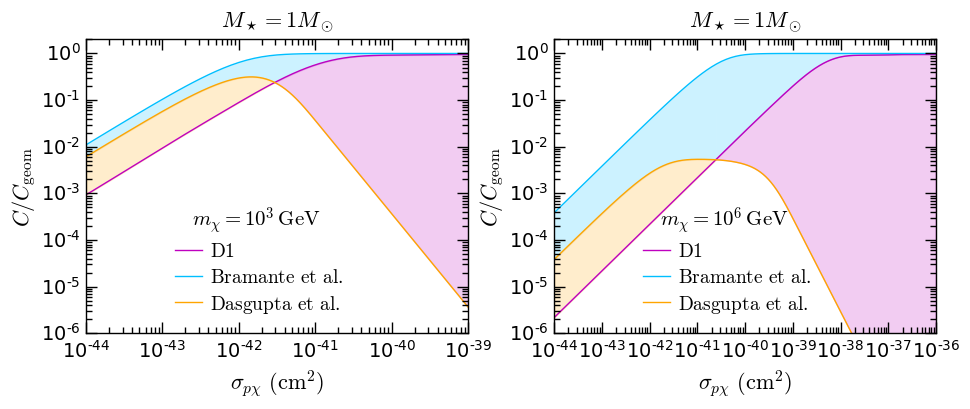}    
    \caption{Capture rate in the multiple scattering regime (magenta) for a $1\Msun$ WD made of carbon, considering scalar-scalar interactions (D1) and $m_\chi=10^3\GeV$ and $10^6\GeV$. 
    We also show results obtained using the prescriptions in refs.~\cite{Bramante:2017xlb} (light blue) and \cite{Dasgupta:2019juq} (orange).\protect\footnotemark }
    \label{fig:Cmulticomparison}
\end{figure}

In Fig.~\ref{fig:Cmulticomparison}, we compare our approach (magenta) 
with previous calculations in the literature that made use of simplifying assumptions, namely those of ref.~\cite{Bramante:2017xlb} (light blue) and ref.~\cite{Dasgupta:2019juq} (orange).
We consider scalar-scalar interactions of $10^3\GeV$ and $10^6\GeV$ mass DM, for a $1\Msun$ WD made of carbon,  located in the solar neighborhood, i.e., we assume $\rho_\chi=0.4\GeV/\cm^3$,  $\vstar=220\km/\s$ and $v_d=270\km/\s$. Radial profiles for the carbon number density and escape velocity for this WD were obtained in ref.~\cite{Bell:2021fye}, 
Our results are given in terms of the maximal capture rate that can be achieved, the so-called geometric limit, which is given by
  \begin{eqnarray}
 C_\mathrm{geom}&=&\frac{\pi R_\star^2 \rho_\chi}{m_\chi} \int_0^\infty \frac{w^2(\Rstar)}{u_\chi} \fMB(u_\chi)du_\chi,\label{eq:Cgeomdef}\\
 &=&\frac{\pi R_\star^2 \rho_\chi}{3v_\star m_\chi} \left[(3 v_\mathrm{esc}^2(\Rstar)+3 v_\star^2+v_d^2) \erf \left(\sqrt{\frac{3}{2}}\frac{v_\star}{v_d}\right) +\sqrt{\frac{6}{\pi}} \vstar v_d e^{-\frac{3\vstar^2}{2 v_d^2}}\right].  \label{eq:Cgeom}
\end{eqnarray}

   \footnotetext{
   Our orange lines were plotted using the definition of the optical depth given in ref.~\cite{Dasgupta:2019juq}. Correcting this definition to impose a cutoff on $\tau$~\cite{Ray:2022oml} would cause the orange lines to plateau to a constant at large cross section (equal to the maximum they reach in Fig.~\ref{fig:Cmulticomparison}) rather than fall off. 
   However, they would not plateau to the geometric capture rate, as the analysis of ref.~\cite{Dasgupta:2019juq} does not treat the large cross section region correctly. For large DM masses, orders of magnitude differences remain.}

As can be seen from Fig.~\ref{fig:Cmulticomparison}, approaches that do not incorporate the radial dependence of the escape velocity, the ionic target number density and form factors, as well as a DM-target relative velocity distribution, overestimate the capture rate by at least one order of magnitude (orange lines) for DM-nucleon cross sections  $\sigma_{p\chi}\lesssim10^{-42}\cm^2$. We note that ref.~\cite{Dasgupta:2019juq} assumed that the nuclear form factor saturates to $\sim0.3$, while ref.~\cite{Bramante:2017xlb} considered $0.5$, which we find is the main cause of the large discrepancy for small cross sections. Recall that we have retained the full radial dependence of the form factors and recoil energies, rather than averaging them over the NS radius. We find that $F_{i,j}^{(N,N')}$ saturates to  $\sim0.034$,  for the scalar-scalar operator and recoil energies lower than  order MeV. 
Note also that the approach of ref.~\cite{Dasgupta:2019juq} does not reach the geometric limit for large cross sections; this occurs because they 
truncate the sum over $C_N$ at a certain maximum value of $N$,  $N_\mathrm{max}$, 
which neglects important terms when the cross section is sufficiently large. 
There is another substantial difference when comparing our results with those calculated using the prescription of ref.~\cite{Bramante:2017xlb} (light blue lines), which neglects the motion of the WD. That approach overestimates both the capture rate and the threshold cross section at which the capture rate saturates the geometric limit.

Regarding the remaining operators in Table~\ref{tab:operatorsle}, operators D2 and D10 are momentum suppressed in the non-relativistic limit, meaning that their matrix elements are proportional to the squared momentum transfer $q_\mathrm{tr}^2$. Therefore, the approximation in Eq.~\ref{eq:diffxsecprop} does not hold. Following the procedure outlined above for non-suppressed interactions, we find that the response function takes the form of a linear combination of hypergeometric functions. For D6, whose cross section also depends on the DM velocity, we can still use the response function in Eq.~\ref{eq:respfuncexp}, but in this case the value of $E_0$ will depend on $u_\chi$.

\subsection{Multiple scattering with multiple targets}
\label{sec:multtarg}

The response function obtained in the previous section is valid for white dwarfs made of a single ionic species. However, the core of a real WDs is composed mainly of two elements plus small traces of other elements. 
In this section, we  generalize our previous result to the case of scattering on two ionic species. 
Each target $i$ will have a different optical depth, $\tau^i_\chi$; the total optical depth $\tau_\chi$ is simply the sum of the contributions from each of the target species.

First, as in the previous section, we consider the probability for DM to interact with a target $i$ and lose energy of at least $\delta E$, while travelling a length $d\tau_\chi^i$, starting from a layer in the WD with optical depth $\tau_\chi^i$. This 
is given by the differential element $G(\tau_\chi^i,\delta_i)d\tau_\chi^i$, where 
\begin{equation}
    \delta_i =\frac{\delta E}{E_0^i}, 
\end{equation}
and the energy scale $E_0^i$ depends on the target $i$ and is calculated using Eq.~\ref{eq:E0}. Thus,  the probability to interact and lose the same amount of energy when DM travels a path-length  $\tau_\chi^i$ is simply  the integral of the differential element over the trajectory, i.e. 
\begin{equation}
   {\cal G}(\tau_\chi^i,\delta_i) = \int_0^{\tau_\chi^i} d\tau \, G(\tau,\delta_i). 
    \label{eq:Gbar}
\end{equation}
Note that this probability satisfies all the expected properties:
\begin{eqnarray}
    \lim_{\tau_\chi^i\to 0}{\cal G}(\tau_\chi^i,\delta_i) &=& 0,\\
    \lim_{\tau_\chi^i\to \infty}{\cal G}(\tau_\chi^i,\delta_i) &=&1, \\
    \lim_{\delta_i\to 0}{\cal G}(\tau_\chi^i,\delta_i) &=& 1-e^{-\tau_\chi^i} \, .
\end{eqnarray}
The first property applies when the DM particle travels through a region that does not contain the target $i$. The second limit tells us that by taking a sufficiently large optical depth, it is possible to make the probability to interact and lose an arbitrary amount of energy as close to 1 as desired. Finally, the third property gives us the probability that DM  interacts with the target $i$; this expression takes the expected form, given that $e^{-\tau_\chi^i}$ is the probability of no interaction with the medium.
Note that ${\cal G}(\tau_\chi^i,\delta_i)$, can be rewritten as
\begin{equation}
     {\cal G}(\tau_\chi^i,\delta_i) = \int_{\delta_i}^\infty d\delta \, k(\tau_\chi^i,\delta),   
\end{equation}
where $k(\tau_\chi^i,\delta)$ is a probability density function, obtained  by differentiating Eq.~\ref{eq:Gbar}
\begin{equation}
     \frac{\partial}{\partial \delta_i}{\cal G}(\tau_\chi^i,\delta_i) = - k(\tau_\chi^i,\delta_i).    
\end{equation}

\begin{figure}
    \centering
    \includegraphics[width=\textwidth]{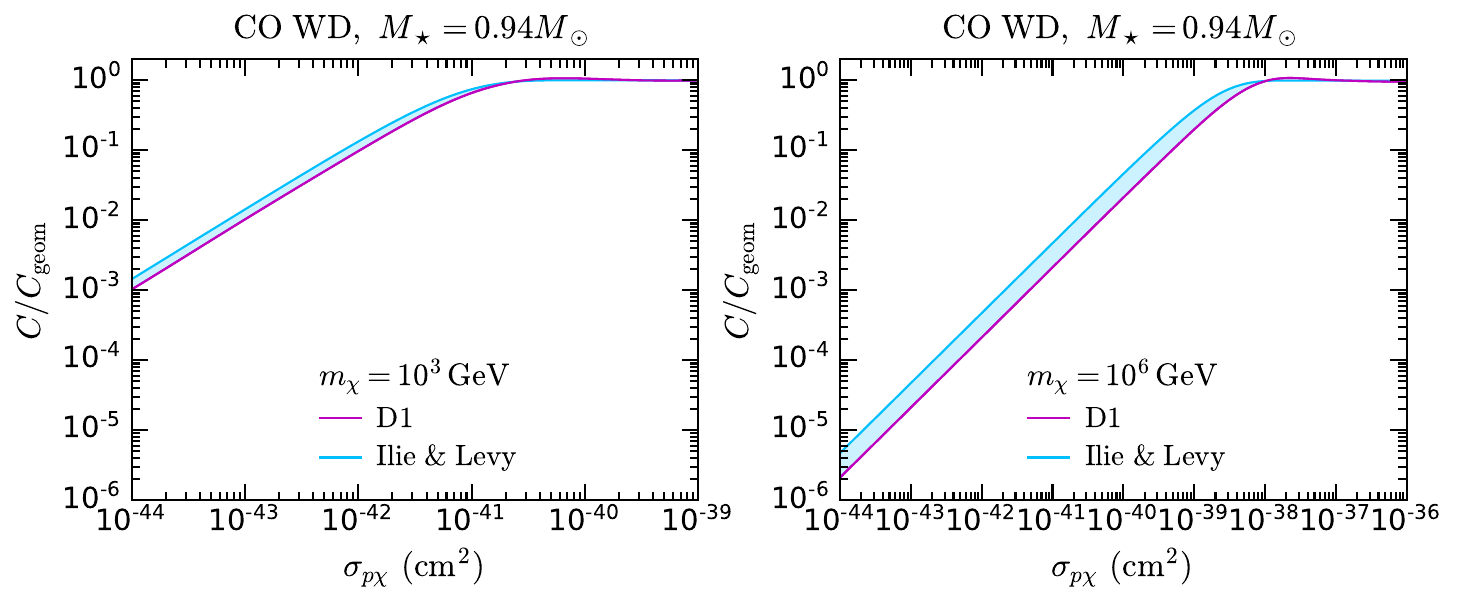}    
    \caption{Capture rate in the multiple-scattering regime (magenta) for a $0.94\Msun$ WD made of carbon and oxygen, for scalar-scalar interactions (D1) and a DM mass of $10^3\GeV$ (left panel)  and $10^6\GeV$ (right panel). 
    We also show results obtained using the prescription in ref.~\cite{Ilie:2021iyh} (light blue).   }
    \label{fig:Cmultitargetcomparison}
\end{figure}

Next, we introduce a second target species. In the presence of these two ionic targets, the cumulative probability of DM to lose an energy $\delta E$ after travelling an optical depth $\tau_\chi^i$ in the target $i$ and $\tau_\chi^j$ in  the second target $j$  is found to be 
\begin{equation}
    {\cal G}_{2,ij}(\delta E) = \int_0^{\delta E/E_0^j} dz \, {\cal G}\left(\tau_\chi^i,\frac{\delta E - z E_0^j}{E_0^i}\right) \left[-\frac{\partial}{\partial z}{\cal G}(\tau_\chi^j,z)\right].
    \label{eq:cumprobG}
\end{equation} 
Since the previous expression is a cumulative probability, 
to calculate the capture rate when the last scattering is over the target $i$, we differentiate Eq.~\ref{eq:cumprobG} with respect to $\tau_\chi^i$ to obtain\footnote{Note that, technically, we should differentiate Eq.~\ref{eq:cumprobG} with respect to $\tau_\chi^j$. However, once we add  $G(\tau_\chi^i,\delta_i)e^{-\tau_\chi^j}$ (which is already differentiated with respect to $\tau_\chi^i$) to Eq.~\ref{eq:ptobG} and arrive at Eq.~\ref{eq:Gmt}, we realize that the integral of the latter is symmetric in the indices $i,j$. Thus,  one can differentiate with respect to either $\tau_\chi^i$ or $\tau_\chi^j$, so we chose the former for convenience.} 
\begin{equation}
     G_{2,ij}(\delta E) = \int_0^{\delta E/E_0^j} dz \, G\left(\tau_\chi^i,\frac{\delta E - z E_0^j}{E_0^i}\right) \left[-\frac{\partial}{\partial z}{\cal G}(\tau_\chi^j,z)\right].   \label{eq:ptobG} 
\end{equation}
This is the probability for DM to be captured between $\tau_\chi^i$ and $\tau_\chi^i+d\tau_\chi^i$.

We are now ready to obtain the capture rate for scattering with two targets. In the previous expression, we have considered that the DM energy is reduced below the threshold for capture to occur after subsequent scatterings with two targets $i$ and $j$; however, capture can happen after collisions with only one ionic species. This is accounted for in the following expression 
\begin{equation}
    G_{12,ij} = G(\tau_\chi^i,\delta_i)e^{-\tau_\chi^j} + G_{2,ij},
    \label{eq:Gmt}
\end{equation}
where the first term represents capture after interactions only with the element $i$. 
As in section~\ref{sec:msrespfunct}, we average Eq.~\ref{eq:Gmt} over the 2 possible trajectories $\tau^-_\chi$ and $\tau^+_\chi$ and the DM angular momentum to obtain the multiple target response function $\tilde{G}_{12,ij}$. 
Thus, the capture rate for the case where scattering from species $i$ dominates is given by
\begin{equation}
   C_i = \frac{\rho_\chi}{m_\chi}\int_0^{R_\star} dr 4\pi r^2 \vesc^2(r) \int_0^\infty du_\chi \frac{\fMB(u_\chi) }{u_\chi} \sum_{i} n_i(r)\sigma_{i\chi}(r)\tilde{G}_{12,ij}\left(r,\frac{m_\chi u_\chi^2}{2 E_0^i}\right).
 \label{eq:CiMT}  
\end{equation}
The contribution of the second target, $C_j$, is obtained by interchanging $i$ with $j$ in Eq.~\ref{eq:CiMT}. Adding the two contributions results in a total capture rate of 
\begin{equation}
   C_{tot} = C_i+ C_j. 
\end{equation}
In principle, this procedure can be generalized to any number of targets.

\begin{figure}[t]
    \centering
    \includegraphics[width=0.5\textwidth]{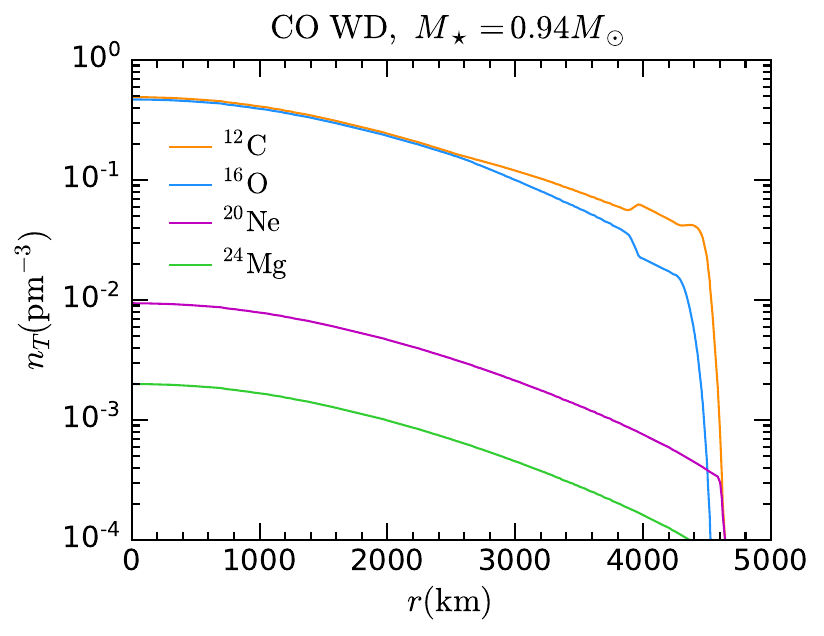}    
    \caption{Radial profiles of the core composition of a $0.94\Msun$ WD simulated with \texttt{MESA}.   
     }
    \label{fig:COWDprofiles}
\end{figure}

In Fig.~\ref{fig:Cmultitargetcomparison}, we compare our approach  with that of ref.~\cite{Ilie:2021iyh} which addresses multi-target multiple-scattering capture in stars, neglecting the stellar structure. We consider a $0.94\Msun$ WD, which is composed mainly of carbon and oxygen. The radial profiles of the constituent number densities are shown in Fig.~\ref{fig:COWDprofiles}. These profiles were obtained by running a simulation of the evolution of a $6\Msun$ main sequence star with solar metallicity, using 
the public stellar evolution code Modules for Experiments in Stellar Astrophysics (\texttt{MESA})~\cite{Paxton:2011,Paxton:2013,Paxton:2015,Paxton:2017eie,Paxton:2019} package, version 21.12.1. 
Note that small traces of elements heavier than oxygen, namely neon and magnesium, are also present in the WD core. 
As in the previous section, we assume the D1 EFT operator, since ref.~\cite{Ilie:2021iyh} only deals with the case of constant DM-nucleon cross section. As we can see, the difference between the two approaches is not as striking as that observed in Fig.~\ref{fig:Cmulticomparison}, with capture rates from ref.~\cite{Ilie:2021iyh} (light blue lines) being at most a factor $\sim2$ larger than our results (magenta lines) for $m_\chi=10^6\GeV$ (right panel). This is due to the fact ref.~\cite{Ilie:2021iyh} adopted the parametrization of the Helm form factor given in ref.~\cite{Lewin:1995rx}, and averaged the form factor and scattering angles for every DM mass considered, which yields more realistic results than those of refs.~\cite{Bramante:2017xlb} and \cite{Dasgupta:2019juq}. Even so, the discrepancy with our results increases with the DM mass.

\begin{figure}[t]
    \centering
    \includegraphics[width=\textwidth]{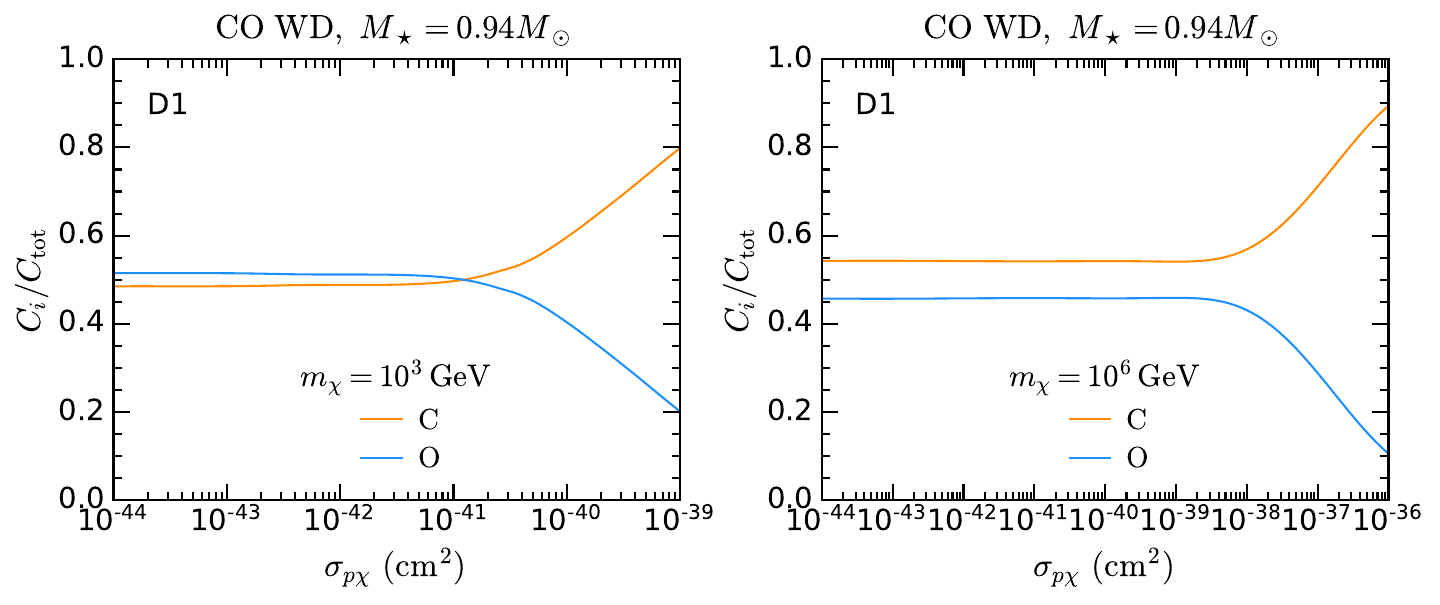}    
    \caption{Contribution of DM scattering with carbon (orange) and oxygen (blue)  targets to the total capture rate in a $0.94\Msun$ WD, for scalar-scalar interactions (D1) and a DM mass of $10^3\GeV$ (left panel)  and $10^6\GeV$ (right panel).
     }
    \label{fig:Cmultitargetcomponents}
\end{figure}

In Fig.~\ref{fig:Cmultitargetcomponents}, we show the contribution of each
WD constituent to the total capture rate for the WD in Fig.~\ref{fig:Cmultitargetcomparison}. We immediately notice that at large cross sections, in the region where the capture rate saturates the geometric limit,  scattering off carbon dominates the capture rate for both DM masses considered. This is due to the fact that in the outer layers of the WD core there are significantly more carbon than oxygen targets (see Fig.~\ref{fig:COWDprofiles}), and in this regime capture occurs closer to the surface. For smaller cross sections,  we observe that 
collisions with oxygen nuclei contribute a significant fraction of the total capture rate, even surpassing the carbon contribution for $m_\chi=10^3\GeV$.  This occurs because, for smaller cross sections, DM can be captured deeper inside the core, where the oxygen capture rate exceeds its carbon counterpart, thus resulting in a more sizeable oxygen contribution. It is also worth noting that oxygen and carbon form factors do not suppress the capture rate to the same degree; the  oxygen form factor gives a larger suppression than that of carbon at high recoil energies, and less at low recoil energies.

\section{Thermalization}
\label{sec:therm}

Once captured, the DM will continue to scatter within the WD, gradually losing energy until it thermalizes in the innermost regions of the star. 
In section~\ref{sec:thermalizationtime}, we outline the calculation of the thermalization timescale for the DM-ion interactions listed in Table~\ref{tab:operatorsle}. As WDs cool, their ions begin to crystallize, forming an ion lattice. The effect of the lattice structure on the thermalization time is discussed in section~\ref{ssec:phonons}.

\subsection{Thermalization time}
\label{sec:thermalizationtime}
The thermalization process can be broadly divided into two stages. The first stage is where the orbit of the captured DM is not wholly contained within the star. The second stage begins once the orbit is fully contained within the star and ends when the DM has reached thermal equilibrium at the WD center. The first stage typically proceeds significantly faster than the second, making up less than 1\% of the total thermalization time~\cite{Kouvaris:2011fi, Garani:2018kkd, Acevedo:2019gre}. Therefore, we focus on the second stage to determine the thermalization time.  
During this last stage, there are two distinctive kinematic regimes depending on whether the DM or target velocity dominates the relative velocity of the interactions. 
The core of a WD is expected to be quite homogeneous in terms of density (see right panel of Fig.~\ref{fig:Tstarevol}) and temperature. As most of the time required to thermalize is confined to the second kinematic regime when DM reaches small energies, we can consider interaction rates in the central region of the WD.

 To estimate the thermalization time, we use the temperature dependent differential interaction rate Eq.~\ref{eq:iondiffintrate}, which accounts for the thermal motion of the nuclei. 
 Summing over the average time between interactions until the energy transferred by the DM reaches the core temperature of the WD~\cite{Bertoni:2013bsa}, $\langle \Delta K_\chi \rangle < \Tstar$, we obtain  
\begin{equation}\label{eq:therm_sum_full}
    \tth = \sum_{n=0}^N \frac{1}{\Omega^-(w(x_n))},
\end{equation}
where $x_n$ is the ratio of the DM kinetic energy $K_\chi$ to the WD core temperature $\Tstar$ after $n$ scatterings, 
and the sum ends for $N$ such that $x_{N}<1$. Each subsequent energy transfer is computed by first calculating the average energy lost in each collision using
\begin{eqnarray}
    \langle \Delta x \rangle &=&  \frac{E(x)}{\Omega^-(w(x))},\\
    E(x) &=& \int_{0}^{x} dy (x-y) R^-(x\rightarrow y),\label{eq:ionenergytransf}
\end{eqnarray} 
where the normalized differential interaction rate (Eq.~\ref{eq:iondiffintrate}) is given in terms of  $x=K_\chi/T_\star$ and represents the probability distribution function for scattering from an initial energy $x\,T_\star$ to a final energy $y\,T_\star$. 
From now on, and in Appendix~\ref{sec:apprate}, we will indicate with $\Omega^-(x)$ the interaction rate as a function of the dimensionless variable $x$, and similarly, we will indicate with $R^-(x\rightarrow y)$ the differential scattering rate from an initial energy $x$ to a final energy $y$.

While the summation in Eq.~\ref{eq:therm_sum_full} can formally be evaluated analytically for some interactions, in general it must be computed numerically. 
When doing so,  numerical instabilities  in the computation of the thermalization time arise from the small nuclear velocities and large DM masses. A sounder approach is then to 
analytically calculate the dominant terms of the interaction rate at high and low (close to thermalization) energy transfers, and use these results to approximate the effects of the thermal motion of the targets.

To obtain analytic expressions for the thermalization time, we can instead approximate the sum in Eq.~\ref{eq:therm_sum_full}  to an integral~\cite{Bertoni:2013bsa} 
\begin{eqnarray}
   \frac{dx}{dt} &=& E(x), \label{eq:thconteq}\\
    \tth &=& \int_{1}^{\infty} \frac{dx}{E(x)} , \label{eq:tthermdef}
\end{eqnarray}
where we have set the initial energy to infinity for simplicity, as the initial energy of the dark matter will be significantly larger than when it is thermalized. 
The validity of this approximation improves for smaller fractional energy losses as an increasing number of collisions are required to thermalize the DM. Specifically, at low energies, we have 
$\langle \Delta x\rangle \approx {\cal O}(1)\times\sqrt{x/\mu}$, while at high energy we have $ \langle \Delta x \rangle \approx 2 x/\mu$;  both approximations are more accurate for large DM masses. 
Note also that we neglect the up-scattering rate. We expect up-scattering to be negligible at large energies and only start to play a role close to thermalization, affecting the result by an $\mathcal{O}(1)$ factor. Hence, all the results presented in this section should be taken as order of magnitude estimates.

To further simplify the analysis, we neglect the effect of the nuclear form factors on the thermalization time. We discuss the effect of including them in Appendix~\ref{sec:formfactors}. In summary, the exponential suppression of the form factors is relevant for energy losses greater than $\langle\Delta K_\chi\rangle \sim 2 K_\chi/\mu \gtrsim E_0\approx \mathcal{O}(\MeV)$. As the thermalization time is dominated by the interactions at lower energies, 
it is not impacted by the form factors.

Eq.~\ref{eq:ionenergytransf} can be evaluated analytically for two distinct scenarios: when $K_\chi\gg T_\star$ (high energy) and when $K_\chi\sim T_\star$ (low energy), which correspond to $x\gg1$  and $x\sim 1$, respectively (see Appendix~\ref{sec:appdeltae}). 
We set $E(x)$ equal to the sum of the high-energy and low-energy contributions to obtain an approximate expression for the energy transfer $E(x)$.  
For differential cross sections proportional to powers of the DM-ion relative velocity  $v_\mathrm{rel}^{2m}$, i.e., cross sections of the form 
\begin{equation}
     \frac{d\sigma_{T \chi }}{d\cos\theta_\mathrm{cm}} = \frac{\sigma_T}{2} v_\mathrm{rel}^{2m}, \label{eq:xsecvrel}
\end{equation}
where $\sigma_T$ is a proportionality constant, 
we obtain 
\begin{equation}
    E(x) \sim  n_T^c\,\sigma_T \,v_T^{2m+1} \sqrt{\frac{x}{\mu}}\left[2\left(\frac{ x}{\mu}\right)^{m+1}+\frac{1}{\sqrt{\pi}}\Gamma\left(m+\frac{3}{2}\right)   \right], 
\end{equation}
where $v_T^2=3\Tstar/m_T$.  
Similarly, for differential cross sections  proportional to powers of the momentum transfer $q_\mathrm{tr}^{2m}$, 
\begin{equation}
    \frac{d\sigma_{T\chi}}{d\cos\theta_\mathrm{cm}} = \frac{\sigma_T(m+1)}{2^{m+1}} \frac{q_\mathrm{tr}^{2m}}{q_0^{2m}},\label{eq:diffxsq0n}
\end{equation}
where $\sigma$ and $q_0$ are constants, 
we find
\begin{equation}
    E(x) \sim  2 n_T^c\,\sigma_T \,v_T \left(\frac{m+1}{m+2} \right)
    \left(\frac{2 v_T^2 m_T^2}{q_0^2}\right)^{m}
    \sqrt{\frac{x}{\mu}}\left[2\left(\frac{ x}{\mu}\right)^{m+1}+\frac{1}{\sqrt{\pi}}\Gamma\left(m+\frac{3}{2}\right)   \right]. 
\end{equation}

  We can now calculate the corresponding thermalization times using Eq.~\ref{eq:tthermdef}; see Appendix~\ref{sec:apptherm}. 
 It is worth noting that the high energy term does not affect this timescale; it is only necessary to ensure that the integral converges at large energies. 
The resulting thermalization time for differential cross sections proportional to powers of the relative speed $v_\mathrm{rel}^{2m}$ is given by
\begin{equation}
    \tth \propto \frac{\mu}{n_T^c\sigma_T v_T } \frac{1}{v_T^{2m}}, \label{eq:tthermvtext}
\end{equation}
where $n_T^c=n_T(0)$ is the ion number density at the WD center. For interactions with cross sections proportional to powers of the momentum transfer $q_\mathrm{tr}^{2m}$, we find that the  thermalization time reads
\begin{equation}
    \tth \propto \frac{\mu}{n_T^c\sigma_T v_T} \left(\frac{q_0^2}{2 v_T^2 m_T^2}\right)^{m}. \label{eq:ttermqtext}
\end{equation}
The precise value of the  $\mathcal{O}(1)$ proportionality constant  in  Eqs.~\ref{eq:tthermvtext} and \ref{eq:ttermqtext} is determined by the coefficients of the expansion of $E(x)$ in $x/\mu$ at all orders, whereas we have only included the highest and lowest order. The exact thermalization time also depends on the up-scattering rate, which we expect to have an impact near the point of thermalization ($x\sim 1$). 
It is worth noting that neglecting the low energy term contribution in $E(x)$,  would have resulted in an incorrect scaling of the thermalization time with respect to $\mu$ at large mass, with a scaling of $\tth \propto \mu^{3/2+m}$ instead of the correct scaling of $\tth \propto \mu$.

To derive $\tth$ for the operators in Table~\ref{tab:operatorsle}, we use the proportionality constants $\sigma_T$ and $q_0$ in Eqs.~\ref{eq:xsecvrel} and \ref{eq:diffxsq0n}, their corresponding thermalization times given by Eqs.~\ref{eq:tthermvtext} and \ref{eq:ttermqtext}, Eq.~\ref{eq:tarnucleonxsec} to relate the DM-nucleus to the DM-nucleon cross section, and the expressions for the differential DM-nucleon cross sections for each operator in the non-relativistic limit given in Appendix~A of ref.~\cite{Bell:2018pkk}. 
We obtain
\begin{eqnarray}
    \tthD{1} &\sim& \frac{\pi  m_\chi m_N^2 \Lambda^4}{n_T^c (c_N^S)^2}\sqrt{\frac{1}{3m_T^9T_\star}}, \label{eq:tthD1}\\
    \tthD{2} &\sim&   \frac{2\pi m_\chi^3 m_N^2 \Lambda^4}{3n_T^c (c_N^S)^2}\sqrt{\frac{1}{3m_T^{11} T_\star^3}},\\
    \tthD{5} &\sim& \frac{\pi m_\chi m_N^2\Lambda^4}{n_T^c (c_N^V)^2}\sqrt{\frac{1}{3m_T^9T_\star}},\\
    \tthD{6} &\sim &\frac{\pi m_\chi m_N^2 \Lambda^4}{2n_T^c (c_N^V)^2}\sqrt{\frac{1}{3m_T^7 T_\star^3}},\\
    \tthD{10} &\sim&\frac{\pi m_\chi m_N^2\Lambda^4}{12\gamma n_T^c (c_N^T)^2}\sqrt{\frac{1}{3m_T^7 T_\star^3}}, \label{eq:tthD10}
\end{eqnarray}
where $\gamma$ in Eq.~\ref{eq:tthD10} is a constant that depends on the specific target (see section \ref{sec:singlescatt}).

\subsection{Effect of the lattice structure}\label{ssec:phonons}

If crystallization has already started  at the very central region of the WD then, because the momentum transfer becomes increasingly smaller during thermalization, the lattice structure of the WD becomes relevant. In this regime, we can no longer assume DM scattering off a free gas of nuclei, and instead must consider the Coulomb lattice of ions as a collective medium. 
This amounts to considering  phonon excitations within the lattice. To account for these effects, we include the dynamic structure function of the lattice $S(q_\mathrm{tr})$ in  the differential cross section 
\begin{equation}
    \frac{d\sigma_{T\chi}}{d\cos\theta_\mathrm{cm}} \rightarrow S(q_\mathrm{tr})\frac{d\sigma_{T\chi}}{d\cos\theta_\mathrm{cm}}. 
\end{equation}
This structure function for a Coulomb lattice is well known~\cite{Baiko:1998xk, Baiko:2000}, and consists of contributions from elastic Bragg scattering and inelastic phonon scattering/absorption. However, only the latter contribute to this current application. For momentum transfers greater than the Brillouin zone, $q_B = (6\pi^2 n_T)^{1/3}$, the structure factor can be approximated as
\begin{equation}
    S(q_\mathrm{tr}) \approx 1 - e^{-2 W(q_\mathrm{tr})},
\end{equation}
where $W(q_\mathrm{tr})$ is the Debye-Waller factor, which for body centered cubic lattices is given by 
\begin{equation}
    W(q_\mathrm{tr})  = \frac{\langle r_T^2\rangle q_\mathrm{tr}^2}{6}, 
\end{equation}
where $\langle r_T^2 \rangle$ is the mean squared separation of ions in the Coulomb lattice. An analytical fit to $W(q_\mathrm{tr})$ with $1\%$ error is given in  ref.~\cite{Baiko:1995} as
\begin{align}
    W(q_\mathrm{tr}) & \simeq  \frac{q_\mathrm{tr}^2}{2m_T\omega_p }\left(1.4 e^{-9.1 \Tstar/\omega_p}+13 \frac{\Tstar}{\omega_p}\right) = \frac{q_\mathrm{tr}^2}{2q_\mathrm{sup}^2},\\
    q_\mathrm{sup}^2 & = \frac{m_T \omega_p}{1.4 e^{-9.1 \Tstar/\omega_p}+13 \Tstar/\omega_p}, \\
    \omega_p & = \sqrt{4\pi Z^2 e^2 n_T/m_T},
\end{align}
where $\omega_p$ is the ion plasma frequency. 

If  $q_\mathrm{tr}^2\ll q_\mathrm{sup}^2$, one can power expand the exponential in the structure factor,  and therefore the interaction rate effectively becomes suppressed by an additional power of $q_\mathrm{tr}^2$. Following a similar treatment to the case of cross sections proportional to $q_\mathrm{tr}^2$, we find that Eqs.~\ref{eq:tthermvtext} and \ref{eq:ttermqtext} now receive an additional suppression factor\footnote{Note that we neglect the additional $\mathcal{O}(1)$ factor $2(m+1)/(m+2)$.} 
\begin{equation}
\frac{q_\mathrm{sup}^2}{6 m_T \Tstar}.
\label{eq:qsupfactor}
\end{equation}
This amounts to increasing the thermalization time by an ${\cal O}(1)$ factor that depends on the equilibrium temperature when accounting for phonon emission. 
Note that this approximation is valid at ${\cal O}(6m_T T/q_\mathrm{sup}^2)$ since we have performed an expansion of the Debye-Waller structure factor. When $6m_T T/q_\mathrm{sup}^2>1$, which occurs for some period of time after the onset of crystallisation in the lighter WDs in Table~\ref{tab:localWDs}, a sound approximation for the thermalization time is obtained by multiplying  Eqs.~\ref{eq:tthermvtext} and \ref{eq:ttermqtext}  by
$\max\left[1,q_\mathrm{sup}^2/(6 m_T \Tstar)\right]$.

Introducing the factor in Eq.~\ref{eq:qsupfactor}  into Eqs.~\ref{eq:tthD1}-\ref{eq:tthD10}, we find the following thermalization times for the SI EFT operators
\begin{eqnarray}
    \tthD{1} &\sim& \frac{\pi  m_\chi m_N^2 \Lambda^4 q_\mathrm{sup}^2}{6n_T^c (c_N^S)^2}\sqrt{\frac{1}{3m_T^{11}T_\star^3}}, \label{eq:tthD1lat}\\
    \tthD{2} &\sim&   \frac{\pi m_\chi^3 m_N^2 \Lambda^4 q_\mathrm{sup}^2}{9n_T^c (c_N^S)^2}\sqrt{\frac{1}{3m_T^{13} T_\star^5}},\\    
    \tthD{5} &\sim& \frac{\pi m_\chi m_N^2\Lambda^4 q_\mathrm{sup}^2}{6n_T^c (c_N^V)^2}\sqrt{\frac{1}{3m_T^{11}T_\star^3}},\\
    \tthD{6} &\sim &\frac{\pi m_\chi m_N^2 \Lambda^4 q_\mathrm{sup}^2}{12n_T^c (c_N^V)^2}\sqrt{\frac{1}{3m_T^9 T_\star^5}},\\
    \tthD{10} &\sim&\frac{\pi m_\chi m_N^2\Lambda^4 q_\mathrm{sup}^2}{72\gamma n_T^c (c_N^T)^2}\sqrt{\frac{1}{3m_T^9 T_\star^5}}. 
\end{eqnarray}

\begin{figure}
    \centering
    \includegraphics[width=0.65\textwidth]{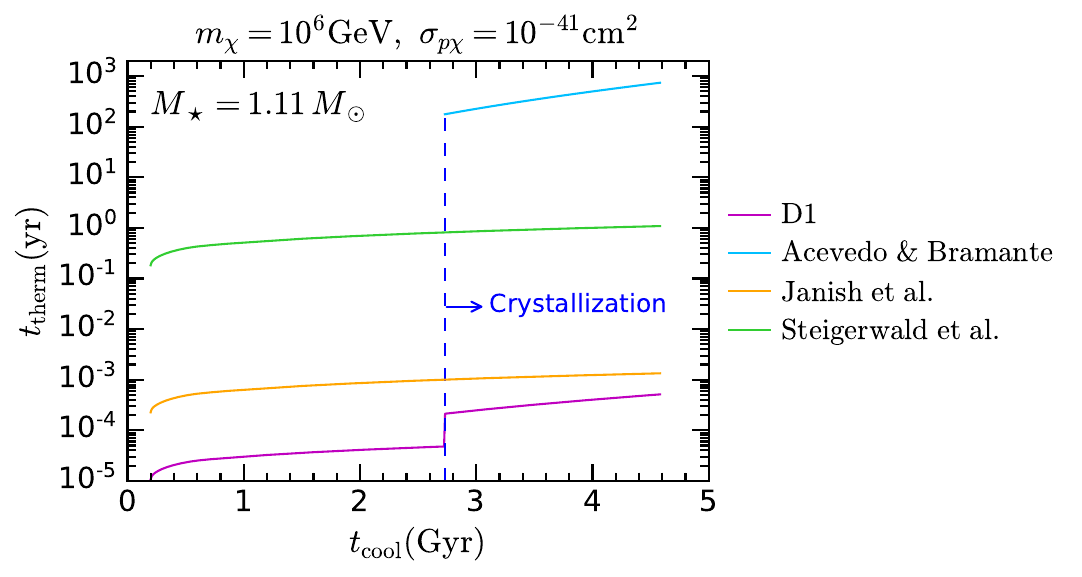}
    \caption{Thermalization time as a function of the WD cooling age for the WD SDSS J232257.27+252807.42 ($\Mstar=1.11\Msun$), DM of mass $m_\chi=10^6\GeV$, and a DM-proton scalar-scalar cross section of $\sigma_{p\chi}=10^{-41}\cm^2$. Our result is depicted in magenta, obtained using the prescription for the second stage of thermalization from ref.~\cite{Acevedo:2019gre} is shown in light blue, while results using the corresponding prescriptions from refs.~\cite{Janish:2019nkk} and \cite{Steigerwald:2022pjo} are depicted in orange and green, respectively.  The dashed blue line represents the onset of crystallisation. }
    \label{fig:thhermcomp}
\end{figure}

Fig.~\ref{fig:thhermcomp} illustrates the variation of the thermalization timescale across the evolution (cooling time $t_{\rm cool}$) of the $1.11\Msun$ WD SDSS J232257.27+252807.42, using the temperature evolution in the left-hand side of Fig.~\ref{fig:Tstarevol}. We consider the scalar-scalar operator (D1), DM of mass $m_\chi=10^6\GeV$ and a DM-proton cross section of $\sigma_{p\chi}=10^{-41}\cm^2$. Before the core of the WD begins  solidifying (at $t_{\rm cool} < 2.73 \Gyr$ for this particular WD; see Table~\ref{tab:localWDs}) we use Eq.~\ref{eq:tthD1}, and after the onset of crystallisation 
we use Eq.~\ref{eq:tthD1lat}. Note the sudden increase in the thermalization time (magenta line) once the crystallisation  front starts moving from the WD center onwards. 

For comparison, we also show in Fig.~\ref{fig:thhermcomp}  results using other prescriptions in the literature. The prescription  for the second stage of thermalization from  ref.~\cite{Acevedo:2019gre}, which  accounts for the effect of the lattice structure so that it holds after the onset of crystallisation, is shown in light blue. This result is orders of magnitude greater than ours. This is mainly due to the fact that this prescription is based on the  expression for the average energy loss at high energy, which, as mentioned in section~\ref{sec:thermalizationtime}, scales as  $\langle\Delta K_\chi\rangle \sim 2 K_\chi/\mu$. Conversely, we find that at the energy relevant for thermalization and, in particular, for phonon emission and absorption (i.e., the low energy transfer regime in which the thermal motion of the targets -- which is included in the scattering rate -- cannot be neglected) the average energy loss is $\langle \Delta K_\chi\rangle \approx {\cal O}(1)\times\sqrt{K_\chi\Tstar/\mu}$. 
Results from the prescriptions in  refs.~\cite{Janish:2019nkk} and \cite{Steigerwald:2022pjo}, which are quite similar and do not consider the lattice structure factor, are depicted in orange and green, respectively.  
Since these two  studies do not account for phonons, comparison with our results should be fair for $t_{\rm cool} < 2.73 \Gyr$ (right hand side of the dashed blue line). 
The main difference between these two calculations is a factor $A^2$ missing in the definition of the DM-target cross section in ref.~\cite{Steigerwald:2022pjo}, which is also one of the main sources of discrepancy with our findings. 
The other main  difference with our results is that both refs.~\cite{Janish:2019nkk} and \cite{Steigerwald:2022pjo} use the DM average energy loss at high energy, thereby neglecting finite temperature effects.

\section{Self-gravitation of asymmetric DM within WDs}
\label{sec:SNignition}

In this section, to showcase the importance of using a proper formalism for capture and thermalization of heavy DM in WDs, we consider the case of asymmetric DM that accumulates in the WD core over time by scattering with the ionic target species. Further collisions, following capture, allow the DM to thermalize to the WD core temperature. In this way, DM settles in the WD interior,  forming a DM sphere close to the center of the star. The general form of the potential energy of this DM cloud, including the WD gravitational potential and self-gravitation, is given by \cite{Steigerwald:2022pjo}
\begin{equation}
 U_{N_\chi} = -4\pi G m_\chi N_\chi(t)\int_0^{\infty} M(r)n_\chi(r) r dr - 4\pi G m_\chi N_\chi(t) \int_0^{\infty} M_\chi(r) n_\chi(r) r dr,
 \label{eq:UN}
\end{equation}
where $N_\chi$ is the number of accumulated DM particles. The quantity $M(r)$ is the WD mass enclosed in a radius $r$, while $M_\chi(r)$ is the mass of DM particles enclosed in the same radius, given by
\begin{equation}
   M_\chi(r)=4\pi m_\chi N_\chi(t)\int_0^r n_\chi(r') r'^2 dr'.  
\end{equation}
The number density of DM particles, $n_\chi$, follows a Maxwell-Boltzmann distribution 
\begin{equation}
   n_\chi(r) = \dfrac{\exp\left[{-m_\chi  \phi(r)/\Tstar}\right]} {\int_0^{\infty} 4\pi r^2 \exp\left[{-m_\chi  \phi(r)/\Tstar}\right]} \simeq \frac{1}{\pi^{3/2} r_\chi^3} \exp{[-r^2/r_\chi^2]},   
   \label{eq:niso}
\end{equation}
where $\phi(r)=-\int_r^\infty G M(r')/r'^2 dr'$ is the gravitational potential and 
\begin{equation}
r_{\chi}=\sqrt{\frac{3 \Tstar}{2\pi G\rho_c m_\chi}} 
\end{equation}
is the scale radius of the DM sphere, which can be obtained using the virial theorem, in the absence of the self-gravitation term.  
In the right-hand side of Eq.~\ref{eq:niso}, we have assumed a constant density in the innermost regions of the WD core \cite{Bell:2021fye}.  
Using this expression in Eq.~\ref{eq:UN}, 
the mean potential energy per DM particle from Eq.~\ref{eq:UN} reads 
\begin{equation}
U =  -2 \pi G \rho_c m_\chi r_\chi^2- \frac{G m_\chi^2 N_\chi(t)}{\sqrt{2\pi} r_\chi},
\label{eq:UDM}
\end{equation}
where $\rho_c=\rho(0)$ is the central density of the WD.

The DM core will grow in mass as the accretion process continues, because asymmetric DM does not self-annihilate. 
When considering the full expression for the potential energy per DM particle (Eq.~\ref{eq:UDM}) in the virial theorem, we find that the necessary condition for collapse is~\footnote{This condition is obtained from the largest real solution of the cubic equation that results from the virial theorem.}  
\begin{equation}
    N_\chi(t) \ge \frac{4\sqrt{2}\pi^{3/2}r_\chi^3\rho_c}{3\sqrt{3} m_\chi} = N_\mathrm{crit}. \label{eq:Ncrit}
\end{equation}
The number of DM particles that have thermalized within the WD center is given by \cite{Steigerwald:2022pjo}
\begin{equation}
    \dfrac{dN_\chi}{dt} = \begin{cases} C(t) \left( 1 + \dfrac{d\tth}{dt}\right)^{-1}, &\qquad t\geq \tth(t) \\
    0, &\qquad t < \tth(t).
    \end{cases}
    \label{eq:dNchidt}
\end{equation}
Integrating this expression, we can calculate the number of  DM particles accumulated within the WD throughout its lifetime. Comparing this result with the collapse condition of Eq.~\ref{eq:Ncrit} allows us to obtain the cutoff scale $\Lambda$ required for collapse and, hence, the corresponding DM-nucleon cross section. 

Note that, in principle, the capture rate $C$ depends on time through the WD temperature (see Fig.~\ref{fig:Tstarevol}). We have checked that for the DM mass range considered here, $C(t)$ can be taken as constant and equal to the value in the $\Tstar\rightarrow0$ approximation derived in section~\ref{sec:capture}. The thermalization time is also a function of the cooling time through the WD core temperature, as shown in Fig.~\ref{fig:thhermcomp}. This time dependence of $\tth$ will be accounted for in our results.

\begin{figure}
    \centering
    \includegraphics[width=0.65\textwidth]{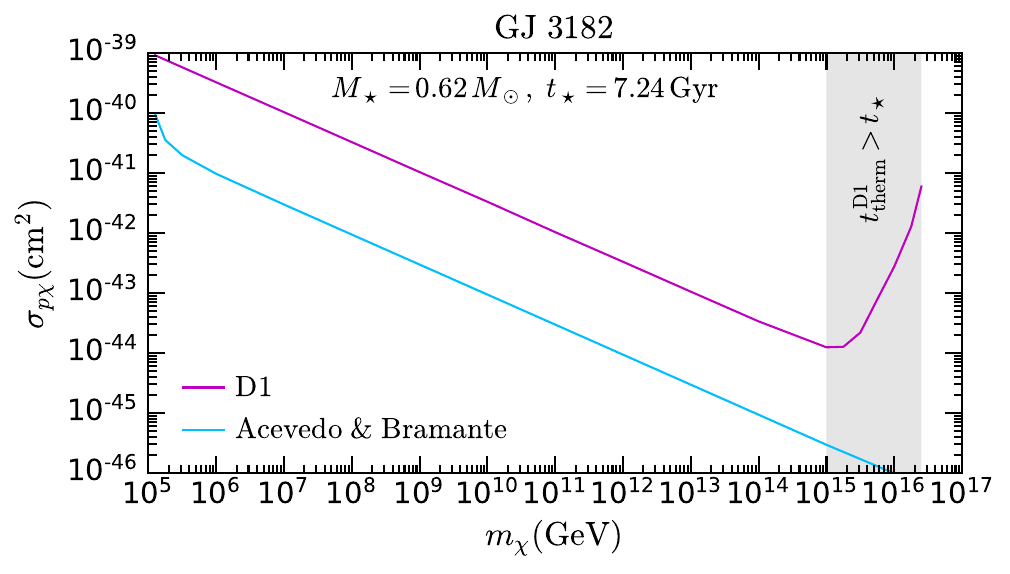}    
    \includegraphics[width=0.65\textwidth]{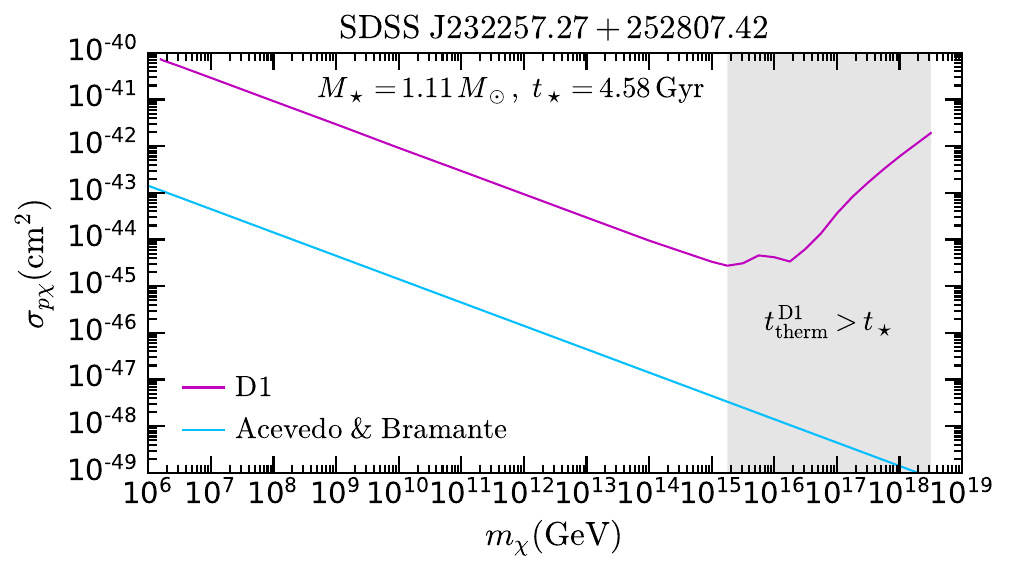}      
    \caption{DM-proton cross section required to reach the self-gravitation condition in Eq.~\ref{eq:Ncrit}, assuming scalar-scalar interactions (D1, in magenta) for the heaviest old WDs in Table~\ref{tab:localWDs}. For comparison we show in light blue results obtained used the condition given in ref.~\cite{Acevedo:2019gre}. 
     }
    \label{fig:zseccritcomp}
\end{figure}

In Fig.~\ref{fig:zseccritcomp}, we show the DM-proton cross section required to reach $N_\mathrm{crit}$ as a function of the DM mass, for two of the local, old WDs in Table~\ref{tab:localWDs}, assuming DM-nucleon scalar-scalar interactions. We arrive at these conclusions by integrating Eq.~\ref{eq:dNchidt}, using the radial profiles for the target number densities obtained in section~\ref{sec:wd} to compute the capture rates given in section~\ref{sec:capture}, together with the thermalization time in Eqs.~\ref{eq:tthD1} or \ref{eq:tthD1lat} using the WD cooling curves given in the left panel of Fig.~\ref{fig:Tstarevol}.

We note that the time derivative of the thermalization timescale in Eq.~\ref{eq:dNchidt}, which is proportional to $m_\chi$,  starts to have an impact for $m_\chi\gtrsim10^{12}\GeV$ ($0.62\Msun$ WD) and $m_\chi\gtrsim10^{11}\GeV$ ($1.11\Msun$ WD). 
As we move to heavier DM masses, we reach a DM mass above which the thermalization is greater than the age of the WD. This is shown by the shaded grey area in Fig.~\ref{fig:zseccritcomp}, and it is responsible for the changes of slope in $\sigma_{p\chi}$. 
We compare these results with the prescription given in ref.~\cite{Acevedo:2019gre} (light blue lines) which neglects the requirement that the DM particles captured in a  time interval $dt$ must thermalize before being added to the DM cloud that has sunk to the center of the WD. The main source of discrepancy with our results, responsible for cross sections that are smaller by $1$ (top panel) and $3$ (bottom panel) orders of magnitude, stems from the computation of the capture rate. This highlights the importance of including the relevant physics of our more complete treatment.  Specifically, including the radial profiles of the escape velocity and target number densities, as well as the nuclear form factors, in the multiple scattering  capture rate, together with a proper estimation of the thermalization time. It is the latter which imposes an upper bound on the mass of non-annihilating DM that can reach self-gravitation in a WD.

\section{Conclusions}
\label{sec:conclusion}

In this paper we have revisited the calculation of the  capture rate of heavy dark matter (DM) in white dwarfs (WDs). In this mass regime, more than one collision is required for the DM to become gravitationally bound to the star. We have extended our formalism for DM capture from the single-scattering to the multiple-scattering regime, by introducing a response function that encodes the cumulative probability for DM to lose an amount of energy of at least $\delta E$ through multiple collisions.
Our treatment incorporates gravitational focusing, nuclear form factors, the variation of the escape velocity along the WD interior and the DM-target relative velocity. We have shown the inclusion of these effects, which are often neglected in the literature, alters the capture rate by orders of magnitude, with the corrections being more critical for very heavy DM.
The response function method is able to handle DM-nucleon interactions that are momentum or velocity suppressed, and can be extended to the case of DM-capture via collisions with multiple targets. It is worth remarking that this formalism is applicable to other stars provided that the dark matter scatters with elements heavier than hydrogen and the escape velocity is much larger than the DM initial velocity far away from the stellar object.

Following capture, the DM will continue to scatter in the star, progressively losing energy until it settles at the center of the star. Eventually, it will reach thermal equilibrium. 
We have estimated the time required for this steady state to be reached, for both a non-crystallized and a crystallized core. In the case of the latter, in-medium effects such as phonon emission and absorption delay thermalization, because the final stages of that process are characterized by low-momentum-transfer DM-target interactions. However, we find that this delay amounts to less than an order of magnitude increase in the thermalization time, which is much smaller than previous estimates.

Finally, to highlight the importance of correctly calculating the multi-scattering capture rate and the thermalization time, we have applied our approach to the case of non-annihilating DM accumulated in a WD core throughout its lifetime. In doing so, we have found orders of magnitude corrections to the DM-nucleon cross sections for which the accumulated DM achieves self-gravitation.

\section*{Acknowledgements}
We thank Maura Ramirez-Quezada for detailed discussions on an earlier version of this manuscript. 
NFB and GB were supported by the Australian Research Council through the ARC Centre of Excellence for Dark Matter Particle Physics, CE200100008.
SR acknowledges partial support from the UK STFC grant ST/T000759/1 and from the Fermi National Accelerator Laboratory (Fermilab), a U.S.
Department of Energy, Office of Science, HEP User Facility. 
MV  was supported by an Australian Government Research Training Program Scholarship. 
SR and MV  thank the  Institute for Nuclear Theory at the University of Washington for its hospitality and the Department of Energy for partial support during the completion of this work.

\appendix

\section{Analytic derivation of the thermalization time}
\label{sec:apprate}

In this section, we derive analytic expressions for the thermalization time for heavy DM. 
In section~\ref{sec:apprates}, we calculate the interaction rate for finite temperature in the high and low energy transfer regimes. In section~\ref{sec:appdeltae}, we estimate the corresponding average energy transfer, and the derivation of the thermalization time can be found in section~\ref{sec:apptherm}. We discuss the effect considering nuclear form factors in the above mentioned calculation in section~\ref{sec:formfactors}.

\subsection{Interaction rate for thermalization}
\label{sec:apprates}

To avoid numerical precision issues, 
it is convenient to obtain an expansion of the interaction rate for low values of $T_\star$. 
To this end, we revisit the expression for the differential interaction rate for finite temperature Eq.~\ref{eq:iondiffintrate}, 
\begin{eqnarray}
   R_T^-(w\rightarrow v) &=&  \int_0^\infty ds  \int_0^\infty dt \,  F(s,t)\frac{4\mu_+^2}{\mu} \frac{n_T(r)v}{w} \frac{d\sigma_{T\chi}}{d\cos\theta_\mathrm{cm}}(s.t,w,v) \Theta(v - |t - s|), \\
   F(s,t) &=& \frac{8\mu_+^2}{\sqrt{\pi}}k^3t\mu \, e^{-k^2u_T^2}\Theta\left(t+s-w\right).
\end{eqnarray} 
 Next, we define  the following functions
\begin{eqnarray}
     \delta_\text{EXP}(x,x_0,c) &=&  c \, e^{-c(x-x_0)}\Theta(x-x_0), \label{eq:deltaexp}\\    
    \delta_\text{G}(x,x_0,c) &=& \frac{c}{\sqrt{\pi}}  e^{-c^2(x-x_0)^2}  \label{eq:deltaG},    
\end{eqnarray} 
where $x$, $x_0$, and $c$ are generic variables. In the limit  $c\rightarrow\infty$, these functions tend to delta functions, i.e.
\begin{align}
    \lim_{c\to\infty}\int_{-\infty}^\infty dx\, \delta_\text{EXP}(x,x_0,c) f(x) &\rightarrow f(x_0),\\
     \lim_{c\to\infty}\int_{-\infty}^\infty dx\, \delta_\text{G}(x,x_0,c) f(x) &\rightarrow f(x_0),
\end{align}
where $f$ is a generic function. 
Using the functions in Eqs.~\ref{eq:deltaexp} and \ref{eq:deltaG}, we rewrite $F(s,t)$ 
\begin{eqnarray}
    F(s,t)\,ds\,dt &=& \delta_\text{EXP}(t^2,(w-s)^2,2\mu\mu_+k^2)\,dt^2 \,\delta_\text{G}\left(s,\frac{\mu w}{2\mu_+},2\mu_+k\right)ds. 
\end{eqnarray}

To express the differential interaction rate in terms of the initial and final kinetic energy of the DM, we make the following substitutions
\begin{align}
    w&=\sqrt{\frac{x}{\mu}}v_T, \qquad
    v=\sqrt{\frac{y}{\mu}}v_T,\\
    s&=\frac{a}{\sqrt{1+\mu}}v_T, \qquad
    t=\frac{b}{\sqrt{1+\mu}}v_T, 
\end{align}
where $v_T^2=3\Tstar/m_T$, $x$ is the ratio of the DM initial kinetic  energy to $T_\star$, which is ${\cal O}(1)$ close to thermalization, and $y$ represents the final DM kinetic energy ratio. This leads to 
\begin{eqnarray}   
    R_T^-(x\rightarrow y) =   \int_{-\infty}^\infty da  \int_{-\infty}^\infty &db&  \,
   F(a,b)\,\frac{2\mu_+^2}{\mu} \frac{n_T(r)}{\sqrt{x}} \frac{v_T}{\sqrt{\mu}} \,
   \Theta\left(\sqrt{y}-\frac{\sqrt{\mu}|a-b|}{\sqrt{1+\mu}}\right) \nonumber\\
    &\times& \frac{d\sigma_{T\chi}}{d\cos\theta_\mathrm{cm}}\left(\frac{a}{\sqrt{1+\mu}}v_T,\frac{b}{\sqrt{1+\mu}}v_T,\frac{\sqrt{x}}{\sqrt{\mu}}v_T,\frac{\sqrt{y}}{\sqrt{\mu}}v_T\right), \label{eq:rate2int}
\end{eqnarray}
where
\begin{equation}
   F(a,b)\,da\,db = \delta_\text{EXP}\left[b^2,\left(\frac{\sqrt{1+\mu}}{\sqrt{\mu}}\sqrt{x}-a\right)^2,\mu\right]db^2 \delta_\text{G}\left(a,\frac{\sqrt{\mu x}}{\sqrt{1+\mu}},\sqrt{1+\mu}\right) da.
\end{equation}

In the large DM mass regime, the third parameter in  $\delta_\text{G}$ and $\delta_\text{EXP}$ in the previous expression is  large since $\mu\gg1$, 
hence  we can expand these functions around the following points
\begin{eqnarray}
    a&=& \frac{\sqrt{\mu x}}{\sqrt{1+\mu}}, \\
    b &=& \frac{\sqrt{1+\mu}}{\sqrt{\mu}}\sqrt{x}-a = \frac{\sqrt{x}}{\sqrt{\mu(1+\mu)}}. 
\end{eqnarray}
Integrating over the delta functions, the resulting expression for the  interaction rate is 
\begin{equation}
    R_T^-(x\rightarrow y) =  \frac{2\mu_+^2}{\mu^{3/2}} \frac{n_T(r) v_T}{\sqrt{x}}   \frac{d\sigma_{T\chi}}{d\cos\theta_\mathrm{cm}}\left(\frac{\sqrt{x}}{\sqrt{\mu}}v_T,\frac{\sqrt{y}}{\sqrt{\mu}}v_T\right) \Theta\left(y-\frac{\mu_{-}^2}{\mu_+^2}x\right). 
    \label{eq:difintratexy}
\end{equation}

We consider  differential cross sections proportional to powers of the DM-ion relative  velocity  $v_\mathrm{rel}^{2m}=w^{2m}$ and to powers of the momentum transfer $q_\mathrm{tr}^{2m}$, namely
\begin{align}
    \frac{d\sigma_{T\chi}}{d\cos\theta_\mathrm{cm}} = \begin{dcases} \dfrac{\sigma_T}{2} w^{2m} = \dfrac{\sigma_T}{2} \left(\frac{x v_T^2}{\mu}\right)^{m}, &\ d\sigma_{T\chi} \propto v_\mathrm{rel}^{2m} \\ 
\dfrac{\sigma_T(m+1)}{2^{m+1} q_0^{2m}} \left( m_\chi^2 \dfrac{w^2-v^2}{\mu} \right)^{m} =\dfrac{\sigma_T(m+1)}{2^{m+1}} \left(\dfrac{m_T^2 v_T^2}{q_0^2}\right)^{m} (x-y)^{m}, &\ d\sigma_{T\chi} \propto q_\mathrm{tr}^{2m} 
   \end{dcases}
\end{align}
where $\sigma_T$ and $q_0$ are normalization constants. 

Plugging the appropriate differential cross section in Eq.~\ref{eq:difintratexy} and integrating over $y$, we recover the interaction rate at high energy.
Thus, for $d\sigma_{T\chi}\propto w^{2m}$ 
\begin{equation}
    \Omega_T^-(x) = \int_0^{x}dy \, R_T^-(x\to y) =  n_T(r)\sigma_T w^{2m+1},
\end{equation}
and for  $d\sigma_{T\chi}\propto q_\mathrm{tr}^{2m}$, we obtain
\begin{equation}
    \Omega_T^-(x)  = n_T(r)\sigma_T w \left[\frac{2 \mu^2 w^2  m_T^2}{(\mu+1)^2 q_0^2}\right]^{m}
    \simeq  n_T(r)\sigma_T w \left(\frac{2  w^2  m_T^2}{q_0^2}\right)^{m}. 
\end{equation}

To obtain the previous results, we have made two assumptions. First, by approximating the  integral over $dt$ to a delta function integral, we are assuming that the DM-target relative speed is equal, in magnitude, to the difference between the center of mass velocity and the DM speed. This approximation only holds when one of the two particles has negligible velocity compared to the other. 
Second,  by approximating  the integral over $s$ to a theta function, we are effectively assuming that for $\mu\gg1$ the relative velocity is equal to the DM speed $w$, thus a negligible target speed is assumed. 
To calculate the interaction rate in the regime where the DM velocity is negligible compared to the target speed, we only make the first assumption and approximate $\delta_\text{EXP}$ to a delta function. Thus, 
we return to Eq.~\ref{eq:rate2int}, and  only integrate over $b$,  so that the interaction rate becomes
\begin{eqnarray}   
    \Omega_T^-(x) &=& \frac{2\mu_+^2}{\mu^{3/2}} \frac{n_T(r) v_T}{\sqrt{x}}   \int_{-\infty}^\infty da  \int_0^{x} dy \, \
   \delta_\text{G}\left(a,\frac{\sqrt{\mu x}}{\sqrt{1+\mu}},\sqrt{1+\mu}\right)
 \Theta\left(\sqrt{y}-\left|2a\frac{\sqrt{\mu}}{\sqrt{1+\mu}}-\sqrt{x}\right|\right) \nonumber\\
    &\times& \frac{d\sigma_{T\chi}}{d\cos\theta_\mathrm{cm}}\left(\frac{a}{\sqrt{1+\mu}}v_T,\frac{\sqrt{x(1+\mu)}-a\sqrt{\mu}}{\sqrt{\mu(1+\mu)}}v_T,\frac{\sqrt{x}}{\sqrt{\mu}}v_T,\frac{\sqrt{y}}{\sqrt{\mu}}v_T\right), 
\end{eqnarray}
where we have swapped the integration order. The theta function sets a lower limit on the integration intervals for $y$ and $a$, which leads to
\begin{eqnarray}   
    \Omega_T^-(x) &=& \frac{2\mu_+^2}{\mu^{3/2}} \frac{n_T(r) v_T}{\sqrt{x}}   \int_{0}^{\sqrt{x(1+\mu)/\mu}} da  \; 
   \delta_\text{G}\left(a,\frac{\sqrt{\mu x}}{\sqrt{1+\mu}},\sqrt{1+\mu}\right) \nonumber \\
   &&\times \int_{y_\mathrm{min}(a)}^{x} dy \, \dfrac{d\sigma_{T\chi}}{d\cos\theta_\mathrm{cm}}\left(\frac{a \, v_T}{\sqrt{1+\mu}},\dfrac{\sqrt{x}}{\sqrt{\mu}}v_T-\dfrac{a \, v_T}{\sqrt{1+\mu}},\dfrac{\sqrt{x}}{\sqrt{\mu}}v_T,\frac{\sqrt{y}}{\sqrt{\mu}}v_T\right)
\end{eqnarray}
where
\begin{equation}
     y_\mathrm{min}(a) = \left(\sqrt{x}-2a\frac{\sqrt{\mu}}{\sqrt{1+\mu}}\right)^2.   
\end{equation} 
Next, we rescale $a\rightarrow (z + \sqrt{\mu x})/\sqrt{1 + \mu}$ for $\delta_G$ to become a pure Gaussian.

As in the high energy transfer regime, we consider cross sections proportional to powers of the DM-target relative velocity which in the low energy regime read 
\begin{eqnarray}
\frac{d\sigma_{T\chi}}{d\cos\theta_{\cm}} \simeq \frac{\sigma_T}{2} {u_T}^{2m} = \frac{\sigma_T}{2}v_T^{2m} z^{2m}, 
\end{eqnarray}
and cross sections that depend on the momentum transfer, 
\begin{eqnarray}
  \frac{d\sigma_{T\chi}}{d\cos\theta_\mathrm{cm}}&=& \frac{\sigma_T(m+1)}{2^{m+1} q_0^{2m}} (2 m_\chi^2 t^2)^m \left[   1-\frac{(s^2+t^2-v^2)(s^2+t^2-w^2)}{4 s^2 t^2}   \right]^{m} \nonumber \\ &=& \frac{\sigma_T(m+1)}{2^{m+1}} \left(\frac{m_T v_T}{q_0}\right)^{2m} \left( \frac{\sqrt{\mu x} - z\mu}{\sqrt{\mu x}+z}\right)^m (x-y)^m. 
\end{eqnarray}

To estimate the interaction rate in this regime, we keep the dominant terms in $\mu$,  for the case of cross sections proportional to $v_\mathrm{rel}^{2m}$, this leads to
\begin{eqnarray}
    \Omega_T^-(x) &\sim&  -\frac{2\mu_+^2}{\mu^{3/2}} \frac{n_T(r) v_T}{\sqrt{x}}\int_{-\sqrt{\mu x}}^{\sqrt{x/\mu}} dz\;  \delta_\text{G}(z,0,1) \frac{4z \sqrt{x}}{\sqrt{\mu}} \frac{\sigma_T}{2}\left(v_T^2z^2\right)^m\\
    &\sim&  \frac{1}{\sqrt{\pi}} n_T(r)\, \sigma_T v_T^{2m+1} \int_0^\infty dz\;  z^{2m+1} e^{-z^2}  = \frac{m!}{2\sqrt{\pi}} n_T(r) \sigma_T v_T^{2m+1}. 
\end{eqnarray}
Similarly, for cross sections that depend on the momentum transfer  $q_\mathrm{tr}^{2m}$, we find 
\begin{eqnarray}
    \Omega_T^-(x) &\sim&  -\frac{2\mu_+^2}{\mu^{3/2}} \frac{n_T(r) v_T}{\sqrt{x}}\int_{-\sqrt{\mu x}}^{\sqrt{x/\mu}} dz\;  \delta_\text{G}(z,0,1) \sigma \left(\frac{m_T}{q_0}\right)^{2m}2^{m+1}z^{2m+1}v_T^{2m}\frac{\sqrt{x}}{\sqrt{\mu}}\\
    &\sim& \frac{m!}{2\sqrt{\pi}} n_T(r)\sigma v_T \left(\frac{2m_T^2 v_T^2}{q_0}\right)^{m}.
\end{eqnarray}

\subsection{Average energy transfers}
\label{sec:appdeltae}

From the previous section,  the energy transfer in units of $T_\star$ is given by $x-y$. Hence, the average energy lost per collision is given by 
\begin{eqnarray}
    \langle \Delta x \rangle = \frac{E(x)}{\Omega^-(w(x))}=  \frac{\int_0^{x}dy \, R_T^-(x\to y)(x-y)}{\int_0^{x}dy \, R_T^-(x\to y)}. 
\end{eqnarray}
As we shall see in the next section,  the computation of the thermalization time only requires evaluating the energy transfer $E(x)$. This can be done in the same manner as for the interaction rate in section~\ref{sec:apprates}. 
In the high energy transfer regime $x\gg1$,  we obtain 
\begin{align}
    E(x) \simeq \begin{dcases} 2n_T(r)\sigma_T  \left(\frac{x }{\mu}\right)^{m+3/2} v_T^{2m+1}, &\quad d\sigma_{T\chi} \propto v_\mathrm{rel}^{2m} \\ 
\frac{4(m+1)}{m+2} n_T(r)\sigma_T  v_T  \left(\frac{x }{\mu}\right)^{m+3/2} \left(\frac{2 m_T^2 v_T^2 }{q_0^2}\right)^{m}, &\quad d\sigma_{T\chi} \propto q_\mathrm{tr}^{2m} 
   \end{dcases}
\end{align}

In the low energy regime, i.e. $x\sim1$ we find 
for cross sections proportional to powers of the DM-ion relative velocity $v_\mathrm{rel}^{2m}$
\begin{equation}
    E(x) \sim \Gamma\left(m+\frac{3}{2}\right)\frac{n_T(r)\sigma_T }{\sqrt{\pi}} \sqrt{\frac{x}{\mu}} v_T^{2m+1},
\end{equation}
while for differential cross sections proportional to $q_\mathrm{tr}^{2m}$ we have 
\begin{equation}
    E(x) \sim \Gamma\left(m+\frac{3}{2}\right)\frac{2(m+1)}{m+2} \frac{n_T(r)\sigma_T v_T}{\sqrt{\pi}} 
    \sqrt{\frac{x}{\mu}}
    \left(\frac{2m_T^2 v_T^2}{q_0^2}\right)^{m}. 
\end{equation}

\subsection{Thermalization times}
\label{sec:apptherm}

Having  calculated the expression for the energy transfer in the high-energy and low-energy regimes, we can now evaluate  Eq.~\ref{eq:tthermdef}. 
The high energy term of $E(x)$ primarily serves as a regulator, allowing us to set the initial temperature to infinity for convenience, which has a negligible impact on the overall result. For cross sections proportional to $v_\mathrm{rel}^{2m}$, we obtain 
\begin{eqnarray}
    E(x) \sim n_T^c\,\sigma_T \,v_T^{2m+1} \sqrt{\frac{x}{\mu}}\left[2\left(\frac{ x}{\mu}\right)^{m+1}+\frac{1}{\sqrt{\pi}}\Gamma\left(\frac{3}{2}+m\right)   \right]. 
\end{eqnarray}
Plugging this result in Eq.~\ref{eq:tthermdef},  we calculate the thermalizaton time 
\begin{eqnarray}
    \tth &=& \int_1^\infty  \frac{dx}{n_T^c\,\sigma_T \,v_T^{2m+1} \sqrt{\frac{x}{\mu}}\left[2\left(\frac{ x}{\mu}\right)^{m+1}+\frac{1}{\sqrt{\pi}}\Gamma\left(\frac{3}{2}+m\right)   \right]}\\
    &\sim& C_m \frac{\mu}{n_T^c\sigma_T v_T } \frac{1}{v_T^{2m}}, \label{eq:tthermv} 
\end{eqnarray}
where $C_m$ is a constant of order $\mathcal{O}(1)$ for $m=0,1,2$ 
\begin{equation}
    C_m = \frac{\pi}{2(m+1)}\csc \left(\frac{\pi}{2m+2}\right) \left[\frac{2\sqrt{\pi}}{\Gamma\left(m+\frac{3}{2}\right)}\right]^{\frac{2m+1}{2m+2}}.
\end{equation}

For differential cross sections proportional to powers of the momentum transfer $q_\mathrm{tr}^{2m}$, we obtain a similar result 
\begin{eqnarray}
    \tth &\sim& C_m \frac{\mu}{n_T^c\sigma_T v_T} \left(\frac{q_0^2}{2 v_T^2 m_T^2}\right)^{m},\label{eq:ttermq}\\
    C_m &=& \frac{\pi(m+2)}{4(m+1)^2}\csc\left( \frac{\pi}{2m+2}\right) \left[\frac{2\sqrt{\pi}}{\Gamma\left(m+\frac{3}{2}\right)}\right]^{\frac{2m+1}{2m+2}}.
\end{eqnarray}

Note that the exact value of the $C_m$ coefficients actually depends on all terms of the series expansion, while here, we have calculated only the dominant terms at high and low energy. Thus, we will consider them to be $\mathcal{O}(1)$ numbers. Here, we have neglected up-scattering, which we expect it to have also an $\mathcal{O}(1)$ effect.

\subsection{Effect of the form factors}
\label{sec:formfactors}

We can determine the effect of the nuclear form factors on the thermalization time by calculating the new value of $E_\text{FF}(x)$ when accounting for them, and comparing it to the expression without them, i.e. 
\begin{eqnarray}
    \frac{E_\text{FF}(x)}{E(x)} &=& \frac{\int_0^x dy (x-y)R_T^-(x\rightarrow y) \exp\left[{-(x-y)\frac{\Tstar}{E_0}}\right]}{\int_0^x dy (x-y)R_T^-(x\rightarrow y)} \\
    &\sim& 1+\mathcal{O}\left(\frac{T_\star}{E_0}\right),
\end{eqnarray}
where $E_0$ is at most  $\mathcal{O}(\MeV)$, their exact value depending on the type of the interaction and the ion target, much greater than the WD core temperature (see Fig.~\ref{fig:Tstarevol}).

As stated in the previous section, the thermalization timescale only depends on the low energy behaviour of the energy transfer, so that when calculating thermalization times, it does not matter that the form factor suppresses the energy transfer at high energies, as this regime does not play any role close to thermalization.




\label{Bibliography}

\lhead{\emph{Bibliography}} 

\bibliography{Bibliography} 

\providecommand{\href}[2]{#2}\begingroup\raggedright\begin{thebibliography}{10}

\bibitem{Bertone:2007ae}
G.~Bertone and M.~Fairbairn, ``{Compact Stars as Dark Matter Probes},''
  \href{http://dx.doi.org/10.1103/PhysRevD.77.043515}{{\em Phys. Rev. D}
  {\bfseries 77} (2008) 043515},
  \href{http://arxiv.org/abs/0709.1485}{{\ttfamily arXiv:0709.1485
  [astro-ph]}}.

\bibitem{McCullough:2010ai}
M.~McCullough and M.~Fairbairn, ``{Capture of Inelastic Dark Matter in White
  Dwarves},'' \href{http://dx.doi.org/10.1103/PhysRevD.81.083520}{{\em Phys.
  Rev. D} {\bfseries 81} (2010) 083520},
  \href{http://arxiv.org/abs/1001.2737}{{\ttfamily arXiv:1001.2737 [hep-ph]}}.

\bibitem{Hooper:2010es}
D.~Hooper, D.~Spolyar, A.~Vallinotto, and N.~Y. Gnedin, ``{Inelastic Dark
  Matter As An Efficient Fuel For Compact Stars},''
  \href{http://dx.doi.org/10.1103/PhysRevD.81.103531}{{\em Phys. Rev. D}
  {\bfseries 81} (2010) 103531},
  \href{http://arxiv.org/abs/1002.0005}{{\ttfamily arXiv:1002.0005 [hep-ph]}}.

\bibitem{Amaro-Seoane:2015uny}
P.~Amaro-Seoane, J.~Casanellas, R.~Sch\"odel, E.~Davidson, and J.~Cuadra,
  ``{Probing dark matter crests with white dwarfs and IMBHs},''
  \href{http://dx.doi.org/10.1093/mnras/stw433}{{\em Mon. Not. Roy. Astron.
  Soc.} {\bfseries 459} no.~1, (2016) 695--700},
  \href{http://arxiv.org/abs/1512.00456}{{\ttfamily arXiv:1512.00456
  [astro-ph.CO]}}.

\bibitem{Dasgupta:2019juq}
B.~Dasgupta, A.~Gupta, and A.~Ray, ``{Dark matter capture in celestial objects:
  Improved treatment of multiple scattering and updated constraints from white
  dwarfs},'' \href{http://dx.doi.org/10.1088/1475-7516/2019/08/018}{{\em JCAP}
  {\bfseries 08} (2019) 018}, \href{http://arxiv.org/abs/1906.04204}{{\ttfamily
  arXiv:1906.04204 [hep-ph]}}.

\bibitem{Panotopoulos:2020kuo}
G.~Panotopoulos and I.~Lopes, ``{Constraints on light dark matter particles
  using white dwarf stars},''
  \href{http://dx.doi.org/10.1142/S0218271820500583}{{\em Int. J. Mod. Phys. D}
  {\bfseries 29} no.~08, (2020) 2050058},
  \href{http://arxiv.org/abs/2005.11563}{{\ttfamily arXiv:2005.11563
  [hep-ph]}}.

\bibitem{Curtin:2020tkm}
D.~Curtin and J.~Setford, ``{Direct Detection of Atomic Dark Matter in White
  Dwarfs},'' \href{http://dx.doi.org/10.1007/JHEP03(2021)166}{{\em JHEP}
  {\bfseries 03} (2021) 166}, \href{http://arxiv.org/abs/2010.00601}{{\ttfamily
  arXiv:2010.00601 [hep-ph]}}.

\bibitem{Bell:2021fye}
N.~F. Bell, G.~Busoni, M.~E. Ramirez-Quezada, S.~Robles, and M.~Virgato,
  ``{Improved treatment of dark matter capture in white dwarfs},''
  \href{http://dx.doi.org/10.1088/1475-7516/2021/10/083}{{\em JCAP} {\bfseries
  10} (2021) 083}, \href{http://arxiv.org/abs/2104.14367}{{\ttfamily
  arXiv:2104.14367 [hep-ph]}}.

\bibitem{Graham:2015apa}
P.~W. Graham, S.~Rajendran, and J.~Varela, ``{Dark Matter Triggers of
  Supernovae},'' \href{http://dx.doi.org/10.1103/PhysRevD.92.063007}{{\em Phys.
  Rev. D} {\bfseries 92} no.~6, (2015) 063007},
  \href{http://arxiv.org/abs/1505.04444}{{\ttfamily arXiv:1505.04444
  [hep-ph]}}.

\bibitem{Bramante:2015cua}
J.~Bramante, ``{Dark matter ignition of type Ia supernovae},''
  \href{http://dx.doi.org/10.1103/PhysRevLett.115.141301}{{\em Phys. Rev.
  Lett.} {\bfseries 115} no.~14, (2015) 141301},
  \href{http://arxiv.org/abs/1505.07464}{{\ttfamily arXiv:1505.07464
  [hep-ph]}}.

\bibitem{Graham:2018efk}
P.~W. Graham, R.~Janish, V.~Narayan, S.~Rajendran, and P.~Riggins, ``{White
  Dwarfs as Dark Matter Detectors},''
  \href{http://dx.doi.org/10.1103/PhysRevD.98.115027}{{\em Phys. Rev. D}
  {\bfseries 98} no.~11, (2018) 115027},
  \href{http://arxiv.org/abs/1805.07381}{{\ttfamily arXiv:1805.07381
  [hep-ph]}}.

\bibitem{Acevedo:2019gre}
J.~F. Acevedo and J.~Bramante, ``{Supernovae Sparked By Dark Matter in White
  Dwarfs},'' \href{http://dx.doi.org/10.1103/PhysRevD.100.043020}{{\em Phys.
  Rev. D} {\bfseries 100} no.~4, (2019) 043020},
  \href{http://arxiv.org/abs/1904.11993}{{\ttfamily arXiv:1904.11993
  [hep-ph]}}.

\bibitem{Janish:2019nkk}
R.~Janish, V.~Narayan, and P.~Riggins, ``{Type Ia supernovae from dark matter
  core collapse},'' \href{http://dx.doi.org/10.1103/PhysRevD.100.035008}{{\em
  Phys. Rev. D} {\bfseries 100} no.~3, (2019) 035008},
  \href{http://arxiv.org/abs/1905.00395}{{\ttfamily arXiv:1905.00395
  [hep-ph]}}.

\bibitem{Steigerwald:2022pjo}
H.~Steigerwald, V.~Marra, and S.~Profumo, ``{Revisiting constraints on
  asymmetric dark matter from collapse in white dwarf stars},''
  \href{http://dx.doi.org/10.1103/PhysRevD.105.083507}{{\em Phys. Rev. D}
  {\bfseries 105} no.~8, (2022) 083507},
  \href{http://arxiv.org/abs/2203.09054}{{\ttfamily arXiv:2203.09054
  [astro-ph.CO]}}.

\bibitem{Bramante:2017xlb}
J.~Bramante, A.~Delgado, and A.~Martin, ``{Multiscatter stellar capture of dark
  matter},'' \href{http://dx.doi.org/10.1103/PhysRevD.96.063002}{{\em Phys.
  Rev.} {\bfseries D96} no.~6, (2017) 063002},
\href{http://arxiv.org/abs/1703.04043}{{\ttfamily arXiv:1703.04043 [hep-ph]}}.

\bibitem{Ilie:2020vec}
C.~Ilie, J.~Pilawa, and S.~Zhang, ``{Comment on \textquotedblleft{}Multiscatter
  stellar capture of dark matter\textquotedblright{}},''
  \href{http://dx.doi.org/10.1103/PhysRevD.102.048301}{{\em Phys. Rev. D}
  {\bfseries 102} no.~4, (2020) 048301},
  \href{http://arxiv.org/abs/2005.05946}{{\ttfamily arXiv:2005.05946
  [astro-ph.CO]}}.

\bibitem{Ilie:2021iyh}
C.~Ilie and C.~Levy, ``{Multicomponent multiscatter capture of dark matter},''
  \href{http://dx.doi.org/10.1103/PhysRevD.104.083033}{{\em Phys. Rev. D}
  {\bfseries 104} no.~8, (2021) 083033},
  \href{http://arxiv.org/abs/2105.09765}{{\ttfamily arXiv:2105.09765
  [astro-ph.CO]}}.

\bibitem{Gould:1991va}
A.~Gould, ``{Big bang archeology: WIMP capture by the earth at finite optical
  depth},'' \href{http://dx.doi.org/10.1086/171057}{{\em Astrophys. J.}
  {\bfseries 387} (1992) 21}.

\bibitem{Bramante:2022pmn}
J.~Bramante, J.~Kumar, G.~Mohlabeng, N.~Raj, and N.~Song, ``{Light dark matter
  accumulating in planets: Nuclear scattering},''
  \href{http://dx.doi.org/10.1103/PhysRevD.108.063022}{{\em Phys. Rev. D}
  {\bfseries 108} no.~6, (2023) 063022},
  \href{http://arxiv.org/abs/2210.01812}{{\ttfamily arXiv:2210.01812
  [hep-ph]}}.

\bibitem{Tolman:1939jz}
R.~C. Tolman, ``{Static solutions of Einstein's field equations for spheres of
  fluid},''
\href{http://dx.doi.org/10.1103/PhysRev.55.364}{{\em Phys. Rev.} {\bfseries 55}
  (1939) 364--373}.

\bibitem{Oppenheimer:1939ne}
J.~R. Oppenheimer and G.~M. Volkoff, ``{On Massive neutron cores},''
\href{http://dx.doi.org/10.1103/PhysRev.55.374}{{\em Phys. Rev.} {\bfseries 55}
  (1939) 374--381}.

\bibitem{Rotondo:2009cr}
M.~Rotondo, J.~A. Rueda, R.~Ruffini, and S.-S. Xue, ``{On the relativistic
  Thomas-Fermi treatment of compressed atoms and compressed nuclear matter
  cores of stellar dimensions},''
  \href{http://dx.doi.org/10.1103/PhysRevC.83.045805}{{\em Phys. Rev. C}
  {\bfseries 83} (2011) 045805},
  \href{http://arxiv.org/abs/0911.4622}{{\ttfamily arXiv:0911.4622
  [astro-ph.SR]}}.

\bibitem{Rotondo:2011zz}
M.~Rotondo, J.~A. Rueda, R.~Ruffini, and S.-S. Xue, ``{The Relativistic
  Feynman-Metropolis-Teller theory for white dwarfs in general relativity},''
  \href{http://dx.doi.org/10.1103/PhysRevD.84.084007}{{\em Phys. Rev. D}
  {\bfseries 84} (2011) 084007},
  \href{http://arxiv.org/abs/1012.0154}{{\ttfamily arXiv:1012.0154
  [astro-ph.SR]}}.

\bibitem{deCarvalho:2013rea}
S.~M. de~Carvalho, M.~Rotondo, J.~A. Rueda, and R.~Ruffini, ``{Relativistic
  Feynman-Metropolis-Teller treatment at finite temperatures},''
  \href{http://dx.doi.org/10.1103/PhysRevC.89.015801}{{\em Int. J. Mod. Phys.
  Conf. Ser.} {\bfseries 23} (2013) 244},
  \href{http://arxiv.org/abs/1312.2434}{{\ttfamily arXiv:1312.2434
  [astro-ph.SR]}}.

\bibitem{Limbach:2022}
M.~A. {Limbach}, A.~{Vanderburg}, K.~B. {Stevenson}, S.~{Blouin}, C.~{Morley},
  J.~{Lustig-Yaeger}, M.~{Soares-Furtado}, and M.~{Janson}, ``{A new method for
  finding nearby white dwarfs exoplanets and detecting biosignatures},''
  \href{http://dx.doi.org/10.1093/mnras/stac2823}{{\em \mnras} {\bfseries 517}
  no.~2, (Dec., 2022) 2622--2638},
  \href{http://arxiv.org/abs/2209.12914}{{\ttfamily arXiv:2209.12914
  [astro-ph.EP]}}.

\bibitem{GaiaDR3:2020}
{\bfseries Gaia} Collaboration, R.~L. {Smart} {\em et~al.}, ``{Gaia Early Data
  Release 3. The Gaia Catalogue of Nearby Stars},''
  \href{http://dx.doi.org/10.1051/0004-6361/202039498}{{\em Astron. Astrphys.}
  {\bfseries 649} (May, 2021) A6},
  \href{http://arxiv.org/abs/2012.02061}{{\ttfamily arXiv:2012.02061
  [astro-ph.SR]}}.

\bibitem{Camisassa:2019}
M.~E. {Camisassa}, L.~G. {Althaus}, A.~H. {C{\'o}rsico}, F.~C. {De
  Ger{\'o}nimo}, M.~M. {Miller Bertolami}, M.~L. {Novarino}, R.~D. {Rohrmann},
  F.~C. {Wachlin}, and E.~{Garc{\'\i}a-Berro}, ``{The evolution of
  ultra-massive white dwarfs},''
  \href{http://dx.doi.org/10.1051/0004-6361/201833822}{{\em Astron. Astrophys.}
  {\bfseries 625} (May, 2019) A87},
  \href{http://arxiv.org/abs/1807.03894}{{\ttfamily arXiv:1807.03894
  [astro-ph.SR]}}.

\bibitem{Bedard:2020}
A.~{B{\'e}dard}, P.~{Bergeron}, P.~{Brassard}, and G.~{Fontaine}, ``{On the
  Spectral Evolution of Hot White Dwarf Stars. I. A Detailed Model Atmosphere
  Analysis of Hot White Dwarfs from SDSS DR12},''
  \href{http://dx.doi.org/10.3847/1538-4357/abafbe}{{\em Astrophys. J.}
  {\bfseries 901} no.~2, (Oct., 2020) 93},
  \href{http://arxiv.org/abs/2008.07469}{{\ttfamily arXiv:2008.07469
  [astro-ph.SR]}}.

\bibitem{York:2000gk}
{\bfseries SDSS} Collaboration, D.~G. York {\em et~al.}, ``{The Sloan Digital
  Sky Survey: Technical Summary},''
  \href{http://dx.doi.org/10.1086/301513}{{\em Astron. J.} {\bfseries 120}
  (2000) 1579--1587}, \href{http://arxiv.org/abs/astro-ph/0006396}{{\ttfamily
  arXiv:astro-ph/0006396}}.

\bibitem{Kleinman:2013}
S.~J. {Kleinman}, S.~O. {Kepler}, D.~{Koester}, I.~{Pelisoli},
  V.~{Pe{\c{c}}anha}, A.~{Nitta}, {\em et~al.}, ``{SDSS DR7 White Dwarf
  Catalog},'' \href{http://dx.doi.org/10.1088/0067-0049/204/1/5}{{\em
  Astrophys. J. Suppl.} {\bfseries 204} no.~1, (Jan., 2013) 5},
  \href{http://arxiv.org/abs/1212.1222}{{\ttfamily arXiv:1212.1222
  [astro-ph.SR]}}.

\bibitem{Kepler:2015}
S.~O. {Kepler}, I.~{Pelisoli}, D.~{Koester}, G.~{Ourique}, S.~J. {Kleinman},
  A.~D. {Romero}, A.~{Nitta}, D.~J. {Eisenstein}, J.~E.~S. {Costa},
  B.~{K{\"u}lebi}, S.~{Jordan}, P.~{Dufour}, P.~{Giommi}, and
  A.~{Rebassa-Mansergas}, ``{New white dwarf stars in the Sloan Digital Sky
  Survey Data Release 10},''
  \href{http://dx.doi.org/10.1093/mnras/stu2388}{{\em \mnras} {\bfseries 446}
  no.~4, (Feb., 2015) 4078--4087},
  \href{http://arxiv.org/abs/1411.4149}{{\ttfamily arXiv:1411.4149
  [astro-ph.SR]}}.

\bibitem{Kepler:2019}
S.~O. {Kepler}, I.~{Pelisoli}, D.~{Koester}, N.~{Reindl}, S.~{Geier}, A.~D.
  {Romero}, G.~{Ourique}, C.~d.~P. {Oliveira}, and L.~A. {Amaral}, ``{White
  dwarf and subdwarf stars in the Sloan Digital Sky Survey Data Release 14},''
  \href{http://dx.doi.org/10.1093/mnras/stz960}{{\em MNRAS} {\bfseries 486}
  no.~2, (June, 2019) 2169--2183},
  \href{http://arxiv.org/abs/1904.01626}{{\ttfamily arXiv:1904.01626
  [astro-ph.SR]}}.

\bibitem{Busoni:2017mhe}
G.~Busoni, A.~De~Simone, P.~Scott, and A.~C. Vincent, ``{Evaporation and
  scattering of momentum- and velocity-dependent dark matter in the Sun},''
  \href{http://dx.doi.org/10.1088/1475-7516/2017/10/037}{{\em JCAP} {\bfseries
  1710} no.~10, (2017) 037},
\href{http://arxiv.org/abs/1703.07784}{{\ttfamily arXiv:1703.07784 [hep-ph]}}.

\bibitem{DelNobile:2013sia}
M.~Cirelli, E.~Del~Nobile, and P.~Panci, ``{Tools for model-independent bounds
  in direct dark matter searches},''
  \href{http://dx.doi.org/10.1088/1475-7516/2013/10/019}{{\em JCAP} {\bfseries
  1310} (2013) 019},
\href{http://arxiv.org/abs/1307.5955}{{\ttfamily arXiv:1307.5955 [hep-ph]}}.

\bibitem{Catena:2015uha}
R.~Catena and B.~Schwabe, ``{Form factors for dark matter capture by the Sun in
  effective theories},''
  \href{http://dx.doi.org/10.1088/1475-7516/2015/04/042}{{\em JCAP} {\bfseries
  04} (2015) 042}, \href{http://arxiv.org/abs/1501.03729}{{\ttfamily
  arXiv:1501.03729 [hep-ph]}}.

\bibitem{Fitzpatrick:2012ix}
A.~L. Fitzpatrick, W.~Haxton, E.~Katz, N.~Lubbers, and Y.~Xu, ``{The Effective
  Field Theory of Dark Matter Direct Detection},''
  \href{http://dx.doi.org/10.1088/1475-7516/2013/02/004}{{\em JCAP} {\bfseries
  1302} (2013) 004},
\href{http://arxiv.org/abs/1203.3542}{{\ttfamily arXiv:1203.3542 [hep-ph]}}.

\bibitem{Gould:1987ir}
A.~Gould, ``{Resonant Enhancements in WIMP Capture by the Earth},''
\href{http://dx.doi.org/10.1086/165653}{{\em Astrophys. J.} {\bfseries 321}
  (1987) 571}.

\bibitem{Bell:2020jou}
N.~F. Bell, G.~Busoni, S.~Robles, and M.~Virgato, ``{Improved Treatment of Dark
  Matter Capture in Neutron Stars},''
  \href{http://dx.doi.org/10.1088/1475-7516/2020/09/028}{{\em JCAP} {\bfseries
  09} (2020) 028}, \href{http://arxiv.org/abs/2004.14888}{{\ttfamily
  arXiv:2004.14888 [hep-ph]}}.

\bibitem{Ray:2022oml}
A.~Ray, {\em {Unravelling the Mystery of Dark Matter with Stars \& Black
  Holes}}.
\newblock PhD thesis, Tata Institute of Fundamental Research, Department of
  Theoretical Physics (DTP), India, Tata Institute of Fundamental Research,
  2022.

\bibitem{Paxton:2011}
B.~{Paxton}, L.~{Bildsten}, A.~{Dotter}, F.~{Herwig}, P.~{Lesaffre}, and
  F.~{Timmes}, ``{Modules for Experiments in Stellar Astrophysics (MESA)},''
  \href{http://dx.doi.org/10.1088/0067-0049/192/1/3}{{\em Astrophys. J. Suppl.}
  {\bfseries 192} no.~1, (Jan., 2011) 3},
  \href{http://arxiv.org/abs/1009.1622}{{\ttfamily arXiv:1009.1622
  [astro-ph.SR]}}.

\bibitem{Paxton:2013}
B.~{Paxton}, M.~{Cantiello}, P.~{Arras}, L.~{Bildsten}, E.~F. {Brown},
  A.~{Dotter}, C.~{Mankovich}, M.~H. {Montgomery}, D.~{Stello}, F.~X. {Timmes},
  and R.~{Townsend}, ``{Modules for Experiments in Stellar Astrophysics (MESA):
  Planets, Oscillations, Rotation, and Massive Stars},''
  \href{http://dx.doi.org/10.1088/0067-0049/208/1/4}{{\em Astrophys. J. Suppl.}
  {\bfseries 208} no.~1, (Sept., 2013) 4},
  \href{http://arxiv.org/abs/1301.0319}{{\ttfamily arXiv:1301.0319
  [astro-ph.SR]}}.

\bibitem{Paxton:2015}
B.~{Paxton}, P.~{Marchant}, J.~{Schwab}, E.~B. {Bauer}, L.~{Bildsten},
  M.~{Cantiello}, L.~{Dessart}, R.~{Farmer}, H.~{Hu}, N.~{Langer}, R.~H.~D.
  {Townsend}, D.~M. {Townsley}, and F.~X. {Timmes}, ``{Modules for Experiments
  in Stellar Astrophysics (MESA): Binaries, Pulsations, and Explosions},''
  \href{http://dx.doi.org/10.1088/0067-0049/220/1/15}{{\em Astrophys. J.
  Suppl.} {\bfseries 220} no.~1, (Sept., 2015) 15},
  \href{http://arxiv.org/abs/1506.03146}{{\ttfamily arXiv:1506.03146
  [astro-ph.SR]}}.

\bibitem{Paxton:2017eie}
B.~Paxton {\em et~al.}, ``{Modules for Experiments in Stellar Astrophysics
  (MESA): Convective Boundaries, Element Diffusion, and Massive Star
  Explosions},'' \href{http://dx.doi.org/10.3847/1538-4365/aaa5a8}{{\em
  Astrophys. J. Suppl.} {\bfseries 234} no.~2, (2018) 34},
  \href{http://arxiv.org/abs/1710.08424}{{\ttfamily arXiv:1710.08424
  [astro-ph.SR]}}.

\bibitem{Paxton:2019}
B.~{Paxton}, R.~{Smolec}, J.~{Schwab}, A.~{Gautschy}, L.~{Bildsten},
  M.~{Cantiello}, A.~{Dotter}, R.~{Farmer}, J.~A. {Goldberg}, A.~S. {Jermyn},
  S.~M. {Kanbur}, P.~{Marchant}, A.~{Thoul}, R.~H.~D. {Townsend}, W.~M. {Wolf},
  M.~{Zhang}, and F.~X. {Timmes}, ``{Modules for Experiments in Stellar
  Astrophysics (MESA): Pulsating Variable Stars, Rotation, Convective
  Boundaries, and Energy Conservation},''
  \href{http://dx.doi.org/10.3847/1538-4365/ab2241}{{\em Astrophys. J. Suppl.}
  {\bfseries 243} no.~1, (July, 2019) 10},
  \href{http://arxiv.org/abs/1903.01426}{{\ttfamily arXiv:1903.01426
  [astro-ph.SR]}}.

\bibitem{Lewin:1995rx}
J.~D. Lewin and P.~F. Smith, ``{Review of mathematics, numerical factors, and
  corrections for dark matter experiments based on elastic nuclear recoil},''
  \href{http://dx.doi.org/10.1016/S0927-6505(96)00047-3}{{\em Astropart. Phys.}
  {\bfseries 6} (1996) 87--112}.

\bibitem{Kouvaris:2011fi}
C.~Kouvaris and P.~Tinyakov, ``{Excluding Light Asymmetric Bosonic Dark
  Matter},'' \href{http://dx.doi.org/10.1103/PhysRevLett.107.091301}{{\em Phys.
  Rev. Lett.} {\bfseries 107} (2011) 091301},
\href{http://arxiv.org/abs/1104.0382}{{\ttfamily arXiv:1104.0382
  [astro-ph.CO]}}.

\bibitem{Garani:2018kkd}
R.~Garani, Y.~Genolini, and T.~Hambye, ``{New Analysis of Neutron Star
  Constraints on Asymmetric Dark Matter},''
  \href{http://dx.doi.org/10.1088/1475-7516/2019/05/035}{{\em JCAP} {\bfseries
  05} (2019) 035}, \href{http://arxiv.org/abs/1812.08773}{{\ttfamily
  arXiv:1812.08773 [hep-ph]}}.

\bibitem{Bertoni:2013bsa}
B.~Bertoni, A.~E. Nelson, and S.~Reddy, ``{Dark Matter Thermalization in
  Neutron Stars},'' \href{http://dx.doi.org/10.1103/PhysRevD.88.123505}{{\em
  Phys. Rev.} {\bfseries D88} (2013) 123505},
\href{http://arxiv.org/abs/1309.1721}{{\ttfamily arXiv:1309.1721 [hep-ph]}}.

\bibitem{Bell:2018pkk}
N.~F. Bell, G.~Busoni, and S.~Robles, ``{Heating up Neutron Stars with
  Inelastic Dark Matter},''
  \href{http://dx.doi.org/10.1088/1475-7516/2018/09/018}{{\em JCAP} {\bfseries
  1809} no.~09, (2018) 018},
\href{http://arxiv.org/abs/1807.02840}{{\ttfamily arXiv:1807.02840 [hep-ph]}}.

\bibitem{Baiko:1998xk}
D.~A. Baiko, A.~D. Kaminker, A.~Y. Potekhin, and D.~G. Yakovlev, ``{Ion
  structure factors and electron transport in dense Coulomb plasmas},''
  \href{http://dx.doi.org/10.1103/PhysRevLett.81.5556}{{\em Phys. Rev. Lett.}
  {\bfseries 81} (1998) 5556--5559},
  \href{http://arxiv.org/abs/physics/9811052}{{\ttfamily
  arXiv:physics/9811052}}.

\bibitem{Baiko:2000}
D.~A. Baiko, D.~G. Yakovlev, H.~E.~D. Witt, and W.~L. Slattery, ``Coulomb
  crystals in the harmonic lattice approximation,''
  \href{http://dx.doi.org/10.1103/physreve.61.1912}{{\em Physical Review E}
  {\bfseries 61} no.~2, (Feb, 2000) 1912--1919},
  \href{http://arxiv.org/abs/physics/9912048}{{\ttfamily physics/9912048}}.

\bibitem{Baiko:1995}
D.~A. {Baiko} and D.~G. {Yakovlev}, ``{Thermal and electrical conductivities of
  Coulomb crystals in neutron stars and white dwarfs},''
  \href{http://dx.doi.org/10.48550/arXiv.astro-ph/9604164}{{\em Astronomy
  Letters} {\bfseries 21} no.~5, (Sept., 1995) 702--709},
  \href{http://arxiv.org/abs/astro-ph/9604164}{{\ttfamily
  arXiv:astro-ph/9604164 [astro-ph]}}.

\end{thebibliography}\endgroup

\end{document}